\documentclass[preprint,superscriptaddress,nofootnoteinbib]{revtex4-1}
\usepackage{graphicx, hyperref, amsmath, amssymb, slashed, color, esvect, latexsym,float}

\DeclareMathOperator{\gev}{GeV}

\newcommand{\cA}{{\cal A}}
\newcommand{\cL}{{\cal L}}
\newcommand{\cM}{{\cal M}}

\newcommand{\cO}{{\cal O}}

\newcommand{\hR}{\widehat{R} }

\newcommand{\ep}{\epsilon}
\newcommand{\half}{\frac{1}{2}}

\newcommand{\tot}{{\rm tot}}
\newcommand{\stau}{ {\widetilde{\tau}} }
\newcommand{\wino}{ {\widetilde{W}^\pm} }
\newcommand{\hino}{ {\widetilde{H}^\pm} }
\newcommand{\cino}{ \chi^\pm }

\newcommand{\pL}{\left(}
\newcommand{\pR}{\right)}
\newcommand{\bL}{\left[}
\newcommand{\bR}{\right]}
\newcommand{\beq}{\begin{equation}}
\newcommand{\eeq}{\end{equation}}
\newcommand{\bea}{\begin{eqnarray}}
\newcommand{\eea}{\end{eqnarray}}
\newcommand{\Eq}[1]{Eq.~(\ref{#1})}
\newcommand{\Eqs}[2]{Eqs.~(\ref{#1}) and (\ref{#2})}
\newcommand{\Eqm}[2]{Eqs.~(\ref{#1}) through (\ref{#2})}
\newcommand{\Sec}[1]{Sec.~\ref{#1}}

\newcommand{\Secm}[2]{Secs.~\ref{#1} through \ref{#2}}
\newcommand{\Fig}[1]{Fig.~\ref{#1}}
\newcommand{\Figs}[2]{Figs.~\ref{#1} and \ref{#2}}
\newcommand{\Tab}[1]{Tab.~\ref{#1}}

\begin{document}

\title{Inspecting the Higgs for New Weakly Interacting Particles}
\author{Clifford Cheung}
\affiliation{California Institute of Technology, Pasadena, CA 91125}
\author{Samuel D. McDermott}
\affiliation{Michigan Center for Theoretical Physics, Ann Arbor, MI 48109\\~\\~\\}
\author{Kathryn M. Zurek}
\affiliation{Michigan Center for Theoretical Physics, Ann Arbor, MI 48109\\~\\~\\}

\begin{abstract}

We explore new physics scenarios which are optimally probed through precision Higgs measurements rather than direct collider searches.
Such theories consist of additional electroweak charged or singlet states which couple directly to or mix with the Higgs boson; particles of this kind may be weakly constrained by direct limits due to their meager production rates and soft decay products.
 We present a simplified framework which characterizes the effects of these states on Higgs physics by way of tree level mixing (with neutral scalars) and loop level modifications (from electrically charged states), all expressed in terms of three mixing angles and three loop parameters, respectively.  The theory parameters are constrained and in some cases even fixed by ratios of Higgs production and decay rates.  
Our setup is simpler than a general effective operator analysis, in that we discard parameters irrelevant to Higgs observables while retaining complex correlations among measurements that arise due to the underlying mixing and radiative effects.   We show that certain correlated observations are forbidden, {\it e.g.}~a depleted ratio of Higgs production from gluon fusion versus vector boson fusion together with a depleted ratio of Higgs decays to $b \bar b$ versus $WW$.  Moreover, we study the strong correlation between the Higgs decay rate to $\gamma\gamma$ and $WW$ and how it can be violated in the presence of additional electrically charged particles.  Our formalism maps straightforwardly onto a variety of new physics models, such as the NMSSM.  We show, for example, that with a Higgsino of mass $m_{\cino_1} \gtrsim 100 \gev$ and a singlet-Higgs coupling of $\lambda=0.7$, the photon signal strength can deviate from the vector signal strength by up to $\sim 40-60\%$ while depleting the vector signal strength by only $5-15\%$ relative to the Standard Model.

\end{abstract}

\maketitle

\section{Introduction}

The ATLAS \cite{A1}  and CMS \cite{C1}  collaborations have presented definitive evidence for the existence of a new, Higgs-like boson with a mass of order $125\gev$.  At present, observations are broadly consistent with Standard Model (SM) expectations \cite{fit}, particularly in the $WW$ \cite{WW}, $ZZ$ \cite{ZZ}, and $b\bar b$ \cite{bb} decay channels, a modest surplus in the $\gamma \gamma$ \cite{gam} channel notwithstanding.

Meanwhile, physics beyond the SM has yet to appear in dedicated searches conducted at the LHC.  Searches for supersymmetry (SUSY), extra dimensions, technicolor, and other models have all turned up empty-handed, suggesting the possibility that precision Higgs measurements might offer our best handle on new physics in the coming years.   While present observations carry large error bars, experimental precision will improve over time, providing more definitive constraints on deviations from a SM Higgs sector.

What manner of new physics would appear {\it first} in precision Higgs physics {\it  rather than} in direct searches?  Such particles are unlikely to be colored, since strongly interacting particles are produced en masse and typically subject to stringent searches involving jets.  Hence, the new states should carry electroweak charges alone---a scenario notoriously difficult to disentangle, even when the new states are relatively light.  In the context of SUSY, for example,  light  charginos or staus can escape detection without the aid of strongly produced squarks or gluinos.
If the new particles couple directly to or mix with the Higgs, however, then precision Higgs measurements may offer our leading experimental probe.

There are myriad theoretical motivations for new particles which couple directly to the Higgs boson.  Indeed, such states are required to regulate the quadratic divergences of the Higgs in any model that addresses the gauge hierarchy problem.  In many cases, these interactions can substantially modify Higgs boson physics \cite{lambdaSUSY,LLS,GKRS,BH,BDF,BIK,AFW,EY,MR,CFKT,EGMT,Eric,social,PR,spans,BGW,BD,ACG,EGST,
CHYZZ,ACDGGR,AM,ACCG,BM,CNW,BPT,BGHSWZ,JSW,nima,HKMT, ABMF,jack,Kitahara,DLP,sunghoon,AG,staus,Carena:2012xa,D'Agnolo:2012mj}, especially in processes like $h \rightarrow \gamma \gamma$.  On their own, however, new electroweak charged states offer diminishing returns for modifying observables like the diphoton branching fraction, except in extreme regions of parameter space with very large couplings  \cite{staustab,lambdaSUSY,nm1,Reece:2012gi}.

On the other hand, the presence of additional scalars in the Higgs sector can change this picture dramatically because mixing between the scalars introduces important tree level effects.  If new particles mix with the Higgs boson, then these states are scalars in the singlet, doublet, triplet, etc.~representations, and will in general acquire vacuum expectation values (VEVs).  Triplet VEVs are disfavored by precision electroweak measurements, while higher order representations are somewhat contrived.  Thus, the most natural case of study is a Higgs sector comprised of mixed singlets and doublets.  

This sequence of logic leads us to the effective theory which is the focus of the present work: the SM augmented by a scalar singlet and doublet which mix with the Higgs boson, all of which can couple to additional new states of arbitrary electroweak charges.  The purpose of this paper is to describe and quantify the phenomenology of this setup in a general, relatively model independent fashion.  Our central conclusions are as follows:

\begin{itemize}

\item 
Three mixing angles and three loop parameters are sufficient to characterize the span of observable effects on Higgs properties in this wide class of theories.  This framework is substantially simpler than a canonical effective Lagrangian approach, which entails new particles and theory parameters which play no role for precision Higgs physics.

\item Experimental observables such as Higgs production and decay rates can be directly ``inverted'' to determine the theory parameters of this setup.  Concretely, one can employ \Fig{tbtd} to ascertain two mixing angles, \Fig{bonVfig} for the third mixing angle, and \Fig{rgamV} for information about the loop parameters.  Critical to this determination are the ratio of gluon fusion and vector boson fusion production, and the Higgs decay rates to photons and bottom quarks relative to massive electroweak gauge bosons.

\item Our framework is simpler than an effective operator formalism which parameterizes arbitrary couplings between $h$ to SM fields and, crucially, it preserves important correlations among observables that encode the underlying effects of mixing and loops.  For example, \Fig{tbtd} shows how certain combinations of observations cannot occur.  Likewise, we investigate the tight correlation between the Higgs decay rate to $\gamma \gamma$ and $WW$ and how it can be can broken through important loop effects.

\item This framework applies to a broad class of models which include additional scalar singlets and doublets that mix with the Higgs, together with new electrically charged particles.  We discuss our results in the specific context of the NMSSM.

\end{itemize}

The outline of this paper is as follows.  In \Sec{Framework} we define our framework and relate the associated theory parameters to physical quantities.  We discuss the effects of tree level mixing in \Sec{sec:tree} and in the effects of loops in \Sec{sec:loop}.  Finally, in \Sec{sec:NMSSM} we discuss the specific application of our framework to the NMSSM before concluding.

\section{Framework}\label{Framework}

\subsection{Theory Parameterization}

Our theoretical framework assumes that the observed Higgs boson, $h$, is an admixture of the neutral components of two Higgs doublets, $\phi_u$ and $\phi_d$, and a singlet, $\phi_s$, which acquire VEVs such that
\bea
\phi_I &=& v_I + H_I, \quad I = u,d,s,
\eea
where $v_u^2+v_d^2 = v^2 \simeq (246\gev)^2$, while $v_u/v_d =\tan\beta$ and $v_s$ are free parameters.  We define $h$ to be the lightest mass eigenstate, which is a linear combination of the field fluctuations
\bea
h &=& \sum_I P_I H_I \\
P_I &=& (\cos \alpha \cos \gamma, -\sin \alpha \cos \gamma , -\sin \gamma),
\eea
where $P_I$ is, by construction, an orthonormal vector that defines a column of the  scalar mixing matrix.   Here $\alpha$ characterizes the mixing between $\phi_u$ and $\phi_d$, while $\gamma$  parameterizes the amount of mixing into $\phi_s$.    A priori, $\alpha$ and $\gamma$ label arbitrary angles in spherical coordinates, so $\alpha$ and $\gamma$ are periodic over domains of size $2\pi$ and $\pi$, respectively.  As we will see later on, many physical observables will depend on these angles with a higher frequency of periodicity.  For later convenience, we also define a difference angle
\bea
\delta &=& \alpha - \beta + \pi/2,
\eea 
which measures deviations from the SM ``decoupling'' limit, $\delta = 0$.

In the SM, it is well known that the couplings of $h$ to other fields are fixed by low-energy Higgs theorems \cite{LET}.  In particular, starting with the SM action below the electroweak symmetry breaking scale, the leading Higgs couplings are obtained by sending $v \rightarrow v + h$, so that all $h$ couplings go like $ \partial / \partial v$.  In our framework, this generalizes to the statement that $h$ couples proportionally to $ \sum_I P_I \partial / \partial v_I$. For any particles that derive mass from electroweak symmetry breaking, it is convenient to define the dimensionless quantities
\bea
d_i&=& \sum_I P_I \eta_{I,i}  \label{coupling}\\
\eta_{I,i}&=& \frac v { m_i }\frac{\partial  m_i }{\partial v_I}  \label{eta},
\eea
where $i$ labels the massive electroweak gauge bosons or fermions and  $m_i$ denotes the mass of particle $i$. Note that $d_i=1$ in the SM limit because the quarks, leptons and massive electroweak gauge bosons all acquire tree level masses from electroweak symmetry breaking alone.   Meanwhile, since the photon and gluon do not acquire mass from electroweak symmetry breaking, \Eq{coupling} does not apply to them. We will define these radiatively induced couplings shortly.

 For the massive electroweak gauge bosons and up- and down-type quarks, we have
\beq \label{dtreemaster}
\begin{array}{rcl cl} 
d_V&=&\cos \gamma \sin (\beta- \alpha) &=& \cos \gamma \cos \delta\\
d_t&=&\cos \gamma \cos \alpha/\sin \beta &=& \cos \gamma  \cos \delta \pL 1+ \tan \delta \cot \beta \pR
\\ d_b&=&-\cos \gamma \sin \alpha/\cos \beta &=& \cos \gamma  \cos \delta \pL 1 - \tan \delta \tan \beta \pR , 
\end{array}
\eeq
where in our setup we assume that $\phi_u$ couples to up-type quarks, $\phi_d$ couples to down-type quarks and leptons, and $\phi_s$ carries no renormalizable couplings directly to quarks, leptons, or SM gauge bosons.  As is well-known, this restriction on $\phi_u$ and $\phi_d$ couplings provides a convenient way for evading stringent constrains on flavor changing neutral currents.  Such a choice can be straightforwardly enforced by discrete symmetries or holomorphy, in the case of SUSY.

Our framework also accounts for the possibility that the $\phi_I$ can couple to additional particles beyond the SM.  
When $i$ labels such a new state, we have
\begin{align} \label{dloopmaster}
\begin{array}{rcl}
d_i &=& \cos \gamma \Big( \cos \alpha ~\eta_{u,i} -\sin \alpha ~\eta_{d,i} \Big) - \sin \gamma ~\eta_{s,i} \\ &=& \cos \gamma \Big( \sin \pL \beta + \delta \pR \eta_{u,i} + \cos \pL \beta +\delta \pR \eta_{d,i} \Big) - \sin \gamma ~\eta_{s,i} ,
\end{array}
\end{align}
where $\eta_{I,i}$ is taken to be an unknown loop parameter that will be constrained by experiment.   Because $i$ labels a new particle, $d_i$ has no counterpart in the SM, but it can be can be straightforwardly extracted from a given ultraviolet model using \Eqs{coupling}{eta}. Because $\eta_{I,i}$ characterizes the power of $v_I$ with which $m_i$ scales we expect  $|\eta_{I,i}|=1$ in renormalizable theories in which the entirety of $m_i$ derives from electroweak symmetry breaking \cite{Cheung:2011aa}.  Absent fine-tuning between tree level mass contributions and electroweak symmetry breaking contributions to $m_i$, the naive expectation is that in general $|\eta_{I,i}| \lesssim 1$.

For couplings that only arise at loop level, $v_I$ dependence enters the action through particle mass thresholds that influence the running couplings of electromagnetism and the strong interactions.  As in the SM, these effects are what induce the couplings of $h$ to photons and gluons.  For these interactions we define
\bea
d_\gamma &=& \underset{i}{\sum}  \cA_{\gamma,i} d_i \label{dgamma}\\
d_g &=&    \underset{i}{\sum}  \cA_{g,i} d_i  \label{dg},
\eea
where $i$ sums over all particles which acquire mass from electroweak symmetry breaking, including new states.  The dimensionless constants $\cA_{\gamma,i}$ and $\cA_{g,i}$ are defined as
\bea
\cA_{\gamma,i} &=& A_{J(i)} (\tau_i) C_{\gamma, i } / \left(  \underset{j \in {\rm SM}}{\sum} A_{J(j)} (\tau_j) C_{\gamma, j } \right) \label{Agamma}\\
\cA_{g,i} &=& A_{J(i)} (\tau_i) C_{g, i } / \left(  \underset{j \in {\rm SM}}{\sum} A_{J(j)} (\tau_j) C_{g, j } \right) \label{Ag},
\eea
and characterize the relative importance of loop corrections from each particle.
Note that $\sum_{i\in {\rm SM}} \cA_{\gamma,i}=\sum_{i\in {\rm SM}} \cA_{g,i}=1$, which enforces that $d_\gamma = d_g=1$ in the SM limit.
 Here $A_{J(i)}(\tau_i)$ are kinematic functions of the spin ($J(i)=0,1/2,1$)  and the mass ($\tau_i=m_h^2/4m_i^2$) of the particle $i$ in the loop.   The functions $A_{J(i)}(\tau_i)$ asymptote to beta function coefficients $b_0=1/3,b_{1/2}=4/3,b_1=-7$ in the $\tau_i \to 0$ limit of infinitely heavy mass of the loop particle; full expressions of the $A_{J(i)}(\tau_i)$ are given in, \emph{e.g.} \cite{Carena:2012xa}.  Here we have defined $C_{\gamma,i}=N_{c, i} Q_i^2$, for $N_{c,i}$ colors and charge $Q_i$, and $C_{g,i}=\frac32 C_2(r_i)$, for a quadratic Casimir $C_2$ of the color representation $r_i$ \cite{Eric}.   Because the Higgs coupling to photons is dominated by the $W$ boson loop (with subdominant and destructively interfering contributions from top and bottom quark loops) and the Higgs coupling to gluons is dominated by top and bottom quarks, the relevant SM contributions to \Eq{dgamma} and \Eq{dg} come from
 \beq
 \begin{array}{rcl}
 \cA_{\gamma,V} &\simeq& 1.277 - 0.006 i \\
  \cA_{\gamma,t} &\simeq& -0.281+0.001i\\
   \cA_{\gamma,b} &\simeq& 0.004 + 0.005 i\\
     \cA_{g,t} &\simeq& 1.050 + 0.077 i\\
   \cA_{g,b} &\simeq& -0.050-0.077 i \label{Anumerical}
\end{array}
 \eeq
 assuming $m_h = 125.5$ GeV.
 Note that while the $b$ contributions are naively negligible, they grow like $\tan^2 \beta$ relative to the $t$ contributions and must be included.

\subsection{Observables Parameterization}

Next, let us consider the dependence of physical quantities on these theory parameters.
To do so, we define
\bea
R[{\cal O}] &=& {\cal O} / {\cal O}_{\rm SM}
\eea
to be the ratio of a given observable $\cal O$ to its SM value, ${\cal O}_{\rm SM}$. In this notation, we find that the following important partial width ratios go as
\bea
R[\Gamma(h\rightarrow VV)] &=& |d_V|^2\\
R[\Gamma(h\rightarrow bb)] &=& |d_b|^2\\
R[\Gamma(h\rightarrow \gamma\gamma)] &=& |d_\gamma|^2 .
\eea
Note that $R[\Gamma(h\rightarrow \ell\ell)]  = R[\Gamma(h\rightarrow bb)] $ in our framework because $\phi_d$ provides the masses for both the down-type quarks and the leptons. 

Important production cross-section ratios go as
\bea
R[\sigma(gg\rightarrow h)] &=& |d_g|^2 \quad \textrm{(gluon fusion)} \label{gfr}\\
R[\sigma(VV\rightarrow h)] &=& |d_V|^2  \quad \textrm{(vector boson fusion)} \label{vbfr}.
\eea
The cross-section ratio for $Vh$ associated production scales the same as for vector boson fusion (VBF), since both processes involve the Higgs coupling to the massive  electroweak gauge bosons.   As noted earlier, our analysis will not include new strongly interacting particles because such states are likely to be observed first in direct collider searches rather than precision Higgs physics.  Furthermore, such states tend to drive a separation of gluon fusion and vector boson fusion production that is not observed in the data \cite{fit}.  In addition, if one is interested in driving enhancements to $h \rightarrow \gamma \gamma$, many models require large scalar mixing (through $A$ terms) \cite{D'Agnolo:2012mj}, which can in turn induce vacuum instability \cite{Reece:2012gi}.   As a result, the dominant contribution to gluon fusion arises from top and bottom quark loops, so 
\bea
|d_g|^2 &\simeq & \left|  \cA_{g, t} d_t + \cA_{g, b} d_b  \right|^2 \label{dgr}
. \label{dgequalsdt}
\eea  
From \Eq{Anumerical} we see that the $d_t$ contribution is weighted more heavily than $d_b$.  However, as noted earlier, this can be compensated by important $\tan \beta$ effects.  Note that \Eqm{gfr}{dgequalsdt} imply that all relevant production modes go as $R[\sigma(jj\rightarrow h)]\propto \cos^2 \gamma$.

It will be convenient to present our results in terms of signal strength modifiers which are employed by experimentalists.  For the process $jj \rightarrow h \rightarrow ii$, we have 
\bea
R_i^j &\equiv&  R[\sigma(jj \rightarrow h) \times {\rm Br}(h \rightarrow ii)] \\&=& R[\sigma(jj \rightarrow h)/\Gamma_\tot]R[\Gamma(h \rightarrow ii)] \\
&=& \hR^j |d_i|^2,
\eea
where $\hR^j \equiv  R[\sigma(jj \rightarrow h)/\Gamma_\tot]$ is defined as the ratio of the production cross-section ratio to the total width ratio and  $\Gamma_\tot$ is the full width, which varies like
\bea
R[\Gamma_\tot] &=&  \sum_{i}  {\rm Br}(h\rightarrow ii) |d_i |^2 
\label{width}.
\eea
Here ${\rm Br}(h\rightarrow ii)$ denotes the SM Higgs branching fraction of $h\rightarrow ii$, where $i$ runs over all kinematically accessible final states.  For our analysis, we use the branching fractions for $m_h = 125.5$ GeV shown in Table~\ref{brtab}.  In principle, there can exist additional particles beyond the SM to which the Higgs can decay.  Throughout the present analysis, however, we neglect the possibility of such new light states below the Higgs mass threshold, so in \Eq{width} $i$ labels SM particles alone.

\begin{table}[t]
\begin{tabular}{|c||c| c|c| c|c| c|c| }
\hline 
$i$& $b$& $W$& $g$ & $\tau$& $c$& $Z$& $\gamma$ \\ \hline
Br$(h\rightarrow ii)$& $56.9\%$& $22.3\%$ & $8.52\%$& $6.24\%$& $2.87\%$& $2.76\%$& $0.228\%$ \\ \hline
\end{tabular}
\caption{Branching fractions for $h\rightarrow ii$ for $m_h =125.5$ GeV from \cite{Dittmaier:2012vm}.}
\label{brtab}
\end{table}%

Because the SM branching ratios are dominated by tree level decays to massive particles that exclusively couple through the doublet Higgses $\phi_u$ and $\phi_d$, we expect that $R[\Gamma_\tot] \propto \cos^2\gamma$ as long as $\gamma$ is not very large.  However, if $\gamma \simeq \cO(\pm \pi/2)$ the decays will be dominated by decays through the singlet component $\phi_s$ and we instead find $R[\Gamma_\tot]  \propto \sin^2\gamma$.  As argued above, $R[\sigma(jj\rightarrow h)] \propto \cos^2 \gamma$ holds for all production channels of interest so $\hR^j$ is independent of $\gamma$ for most mixing angles, but as $\gamma \to \pm \pi/2$ the loop level contribution to the width from $\phi_s$ begins to dominate the full width, in which case $\hR^j \propto \cot^2 \gamma \to 0$.

\section{Tree Level Effects}\label{sec:tree}

\begin{figure}[h]
\begin{center}
\includegraphics[width=.47\textwidth]{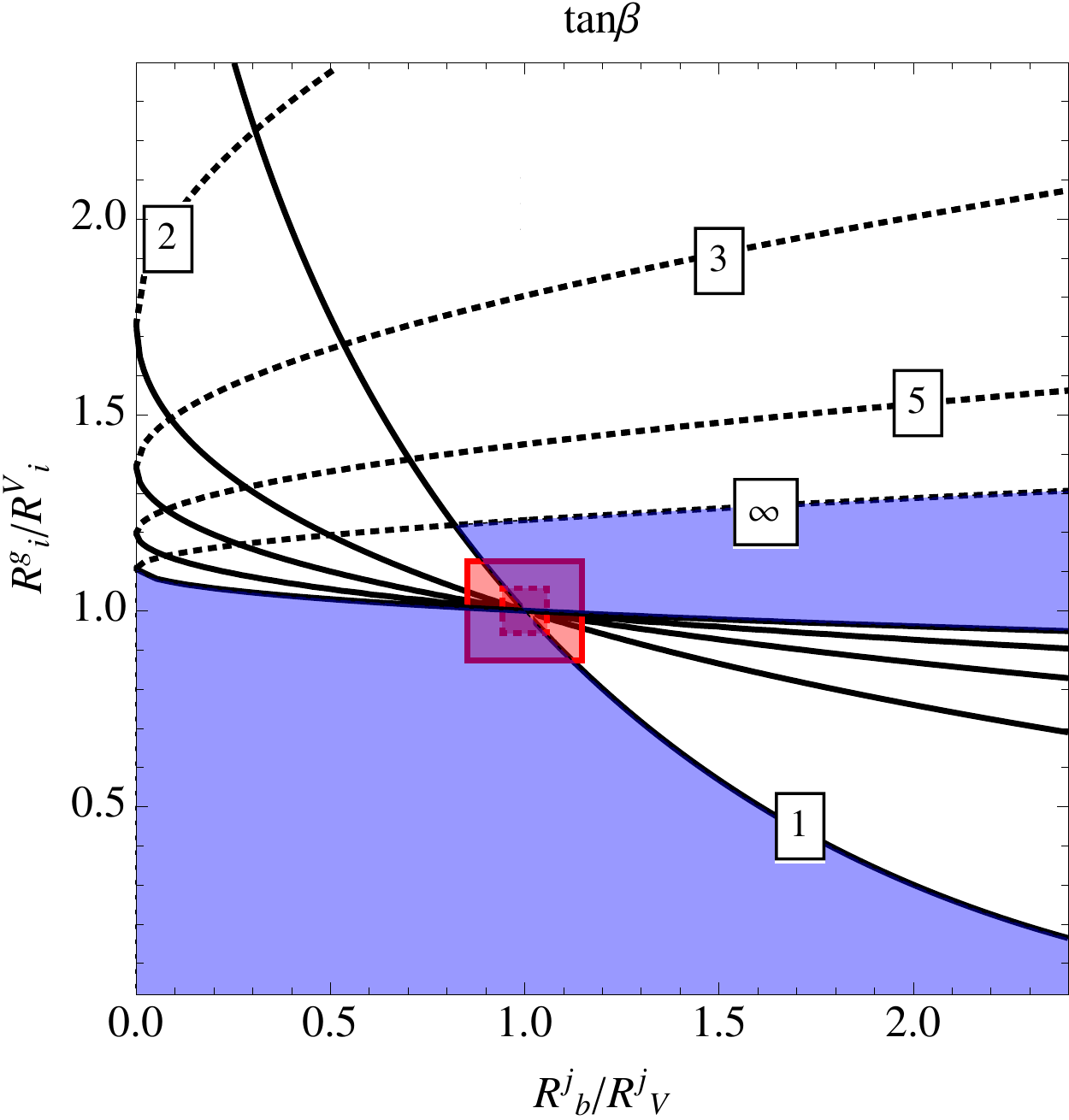}
\includegraphics[width=.47\textwidth]{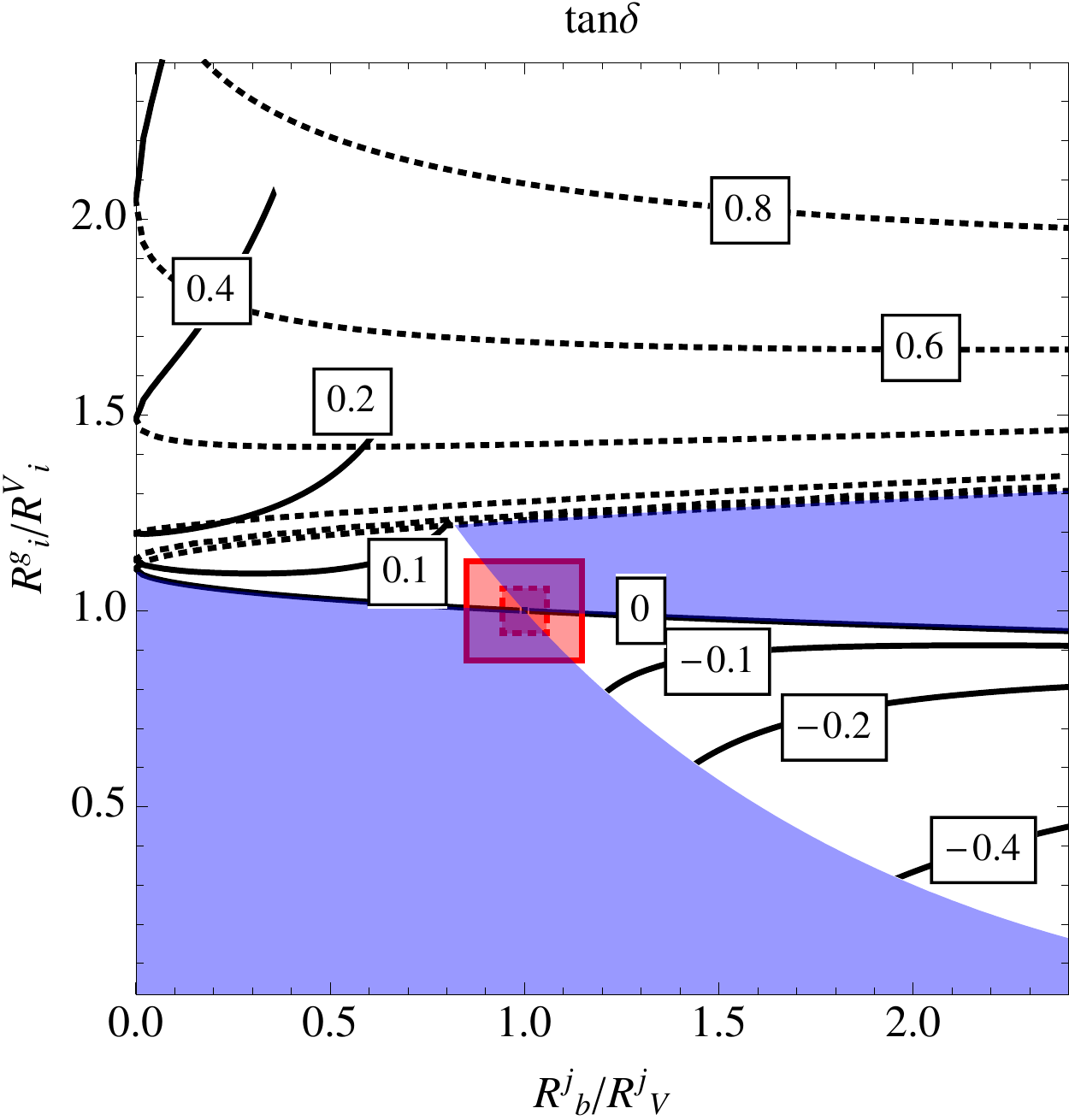}
\caption{Contours of $\tan \beta$ (left) and $\tan \delta$ (right) as functions of $(R^g_i / R^V_i , R^j_b / R^j_V)$, which are obtained directly from experiment. The red shaded regions with solid (dashed) borders show values that will remain consistent with the SM with 300 (3000) ${\rm fb}^{-1}$ and blue shaded regions show values 
which cannot be observed within this framework. 
The solid (dotted) curves show the region where  
$d_b/d_V$ is positive (negative).}
\label{tbtd}
\end{center}
\end{figure}

In this section we consider the effect on Higgs properties from tree level modifications to the SM scalar sector.  It will be particularly convenient to consider certain ratios of quantities in order to reduce systematics, including
\bea
\frac{R_b^j}{R_V^j}&=& \frac{R_\ell^j}{R_V^j} =\frac{|d_b|^2}{|d_V|^2} =
\pL 1 - \tan \delta \tan \beta \pR^2 \label{RboverRV},
\eea
where the production mode $j$ is arbitrary, {\it e.g.}~gluon fusion or vector boson fusion.  We also consider the ratio
\bea
\frac{R_i^g}{R_i^V}&\simeq& \frac{1}{|d_V|^2}  \left|  \cA_{g,t}  d_t + \cA_{g,b}  d_b  \right|^2  =  \left|  \cA_{g,t}  (1+\tan \delta \cot \beta) + \cA_{g,b}  (1-\tan \delta \tan \beta) \right|^2 ,\label{RgoverRV}
\eea
where the decay mode $i$ is arbitrary.
Interestingly, \Eq{RboverRV} and \Eq{RgoverRV} provide two equations for two unknowns, $\delta$ and $\beta$, which can be solved for in terms of experimental inputs. To facilitate this mapping, we plot $\tan \delta$ and $\tan \beta$ as functions of the signal strength modifiers $R^j_b / R^j_V$ and $R^g_i / R^V_i$ in Fig.~\ref{tbtd}. The solid (dotted) curves show the region where  $d_b/d_V$ is positive (negative). For models which are relatively SM-like, this quantity should be positive and close to unity; large new physics contributions are required to flip its sign. The shaded red boxes in Fig.~\ref{tbtd} denote the range of measurements consistent within 1$\sigma$ uncertainty of the SM measured at the LHC at 14 TeV energies and 300 and 3000 ${\rm fb}^{-1}$ luminosity. Any theoretical models enclosed by this region will be very difficult to distinguish from a SM hypothesis. Outside of the red region, however, \Fig{tbtd} can be straightforwardly used to extract the mixing angles.  We list the relative errors that will be compatible with the SM in \Tab{tabunc}, combining in quadrature and making the identification $d_Z = d_V$ from the values given in Table 2.3 of \cite{ATLAS-collaboration:2012iza}.

\begin{table}[t]
\centering
\begin{tabular}{|c||c|c|c|c|c|c|c||c|c|c|c|c|c|c|c|c|}
\hline ratio & $R^g_i/R^V_i$ & $R^j_b/R^j_V$ & $R^j_\gamma/R^j_V$  \\ \hline
uncertainty  & 12.6\% (5.76\%)  & 14.8\% (5.76\%)  & 11.5\% (3.61\%)  \\ \hline
\end{tabular}
\caption{Projected uncertainties on ratios of signal strength modifiers for the 14 TeV LHC with 300 (3000) fb$^{-1}$ of data.}
\label{tabunc}
\end{table}%

The blue regions in \Fig{tbtd} indicate the parameter space of observables which cannot be observed within this framework.  The forbidden regions in the upper right and lower right quadrants cannot occur because $\tan \beta$ falls outside of the allowed range $ \tan\beta >1$.  Far outside of this range, the top quark Yukawa coupling becomes non-perturbative.   Thus, a general prediction is that $R^j_b / R^j_V$ and $R^g_i / R^V_i$ should not be observed deep within these blue regions.  Furthermore, that the lower left quadrant of \Fig{tbtd} is disallowed can be easily understood as follows.  In order to decrease $R^j_b / R^j_V$ and $R^g_i / R^V_i$ simultaneously, one requires a suppression of the couplings of the Higgs to both the top {\it and} bottom quarks relative to the vector bosons.  However, because we consider theories in which $\phi_u$ and $\phi_d$ couple exclusively to up-type and down-type quarks, respectively, the coupling of the Higgs to top and bottom quarks are necessarily anti-correlated.  Similar logic would imply that entirety of the upper right quadrants of \Fig{tbtd} should also be forbidden.  This is true  if $d_b >0$; however, if new physics contributions are so large as to flip the sign to $d_b <0$, then these regions in the upper right quadrants become allowed again.

\begin{figure}[h]
\begin{center}
\includegraphics[width=\textwidth]{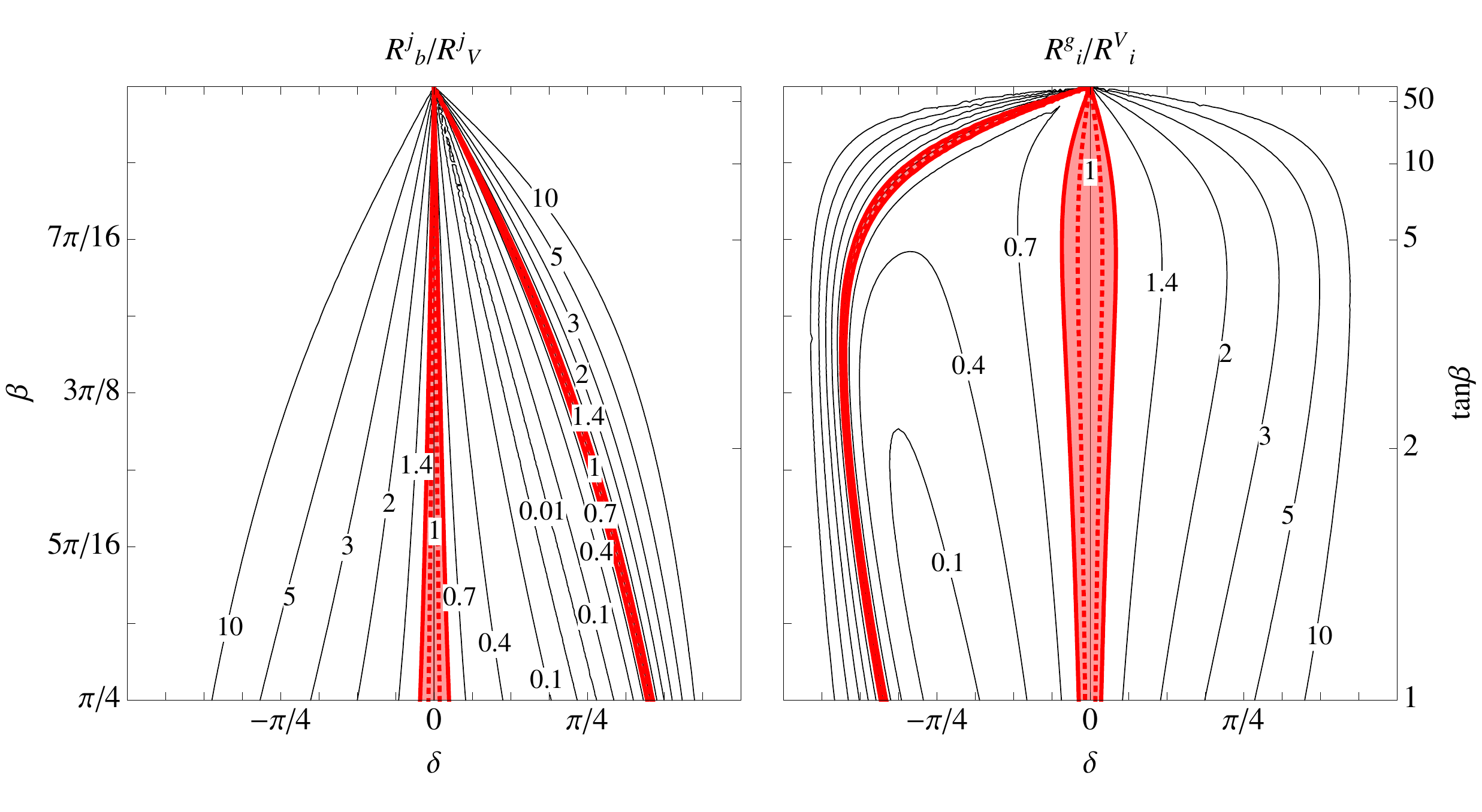}
\caption{ Contours of $R_b^j/R_V^j$ (left) and $R^g_i / R^V_i$ (right) as a function of  theory parameters $(\delta, \beta)$. The red shaded regions are as in \Fig{tbtd}. }
\label{bonVfig}
\end{center}
\end{figure}

\Fig{bonVfig} is a complementary representation of the same information as in \Fig{tbtd}, except that $R^j_b / R^j_V$ and $R^g_i / R^V_i$ are shown as functions of $\tan \delta$ and $\tan \beta$.  One can see that $\tan \delta$ has support in two distinct bands in each plot of \Fig{bonVfig}, but the allowed value is constrained by matching both measurements, as corroborated by \Fig{tbtd}.  One also notices  that $R^j_b / R^j_V$ and $R^g_i / R^V_i$ are quite sensitive to $\delta$, but not so dependent on $\beta$.  For this reason, a precise measurement of $\beta$ is more difficult than a determination of $\delta$, as confirmed by the contours in \Figs{tbtd}{bonVfig}. This is because $\tan \beta$ cannot be extracted in the decoupling limit, so any constraint on $\tan \beta$ necessarily requires an observed deviation in SM-like behavior through $\tan \delta$ first.

\begin{figure}[t]
\begin{center}
\includegraphics[width=6.in]{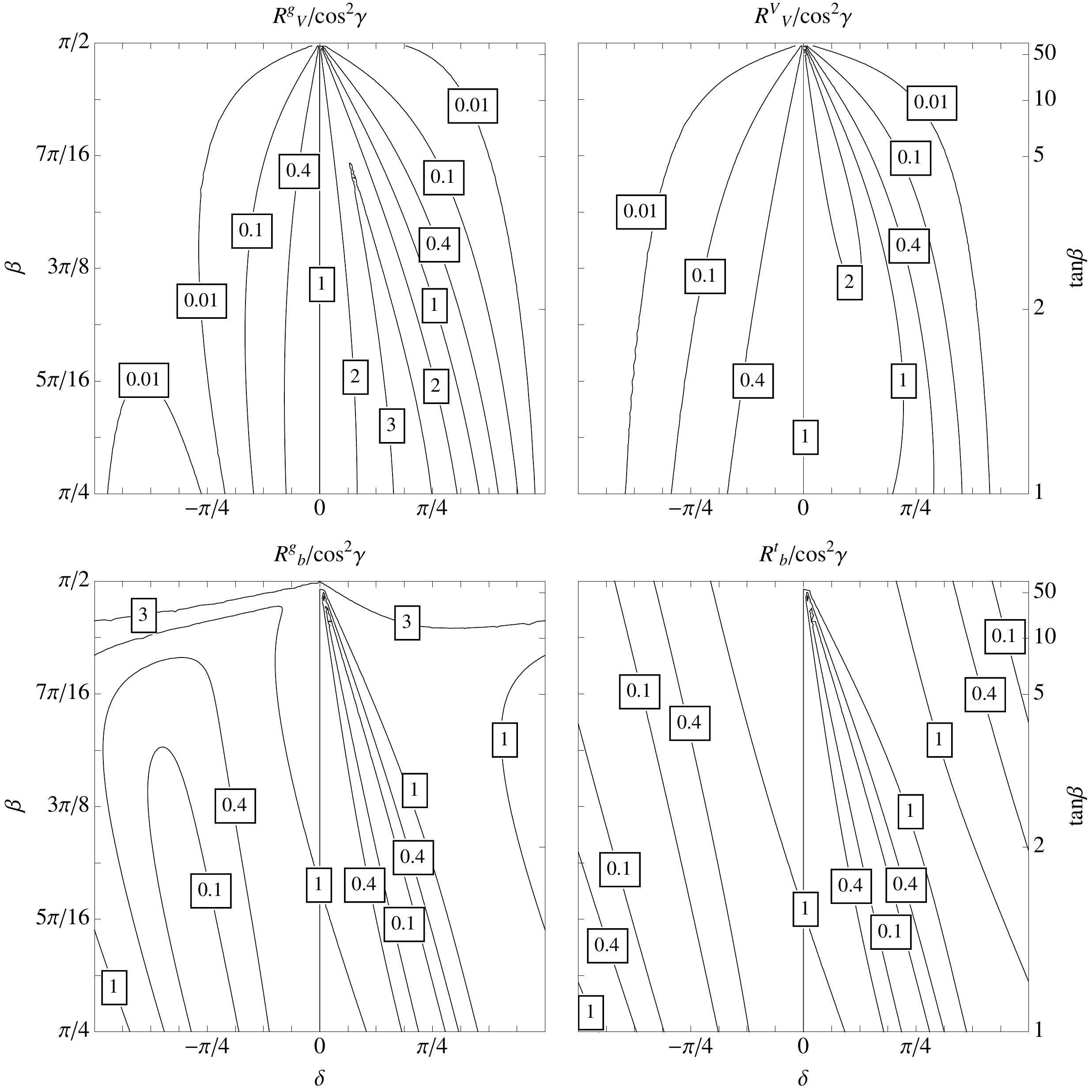}
\caption{Contours of $R^j_i/\cos^2\gamma$ (which we approximate by $R^j_i|_{\gamma=0}$) for $V$ (top row) and $b$ (bottom row) final states from a variety of production channels.  Once $\delta$ and $\beta$ are determined from observation, measuring $R^j_i$ can then be used to obtain $\gamma$. 
}
\label{rjv}
\end{center}
\end{figure}

Once $\delta$ and $\beta$ are determined from data, they can be used to extract other theory parameters from Higgs measurements.  Concretely, given values of $\delta$ and $\beta$, the quantity $\cos^2 \gamma$ can be inferred from a number of different observables, including $R^g_V$, $R^V_V$, $R^g_b$, and $R^t_b$ (where the $t$ superscript denotes top quark associated Higgs production).  Crucially, to a very good approximation these quantities all carry the same $ \cos^2\gamma$ dependence, at least when $\gamma \neq \pm \pi/2$.  Away from that limit of large $\gamma$ we have $R^j_i |_{\gamma=0}  \simeq R^j_i / \cos^2 \gamma$, which we plot as a function of $\beta$ and $\delta$ in \Fig{rjv}.   If $\beta$ and $\delta$ have been extracted from observables, and $R^j_i$ has been measured, then $\gamma$ can be extracted from \Fig{rjv}.

We now explain some of the features of \Fig{rjv}, starting with the upper left panel. By going to the anti-decoupling limit, $\delta = \pm \pi/2$, we can tune the vector coupling arbitrarily low while maintaining a nonzero Higgs width to SM particles. In contrast, it is not possible to increase $R_V^j$ without bound: there is a maximum around where the $b$ width is zero. Taking $\delta>0$ $(\delta < 0)$ corresponds to increasing the amount of $\phi_u$ ($\phi_d$) in the physical Higgs, which boosts the coupling of $h$ to top (bottom) quarks.  Thus, for $\delta >0$ the gluon fusion rate $\sigma(gg\rightarrow h)$ is boosted through the top loop, while the full width $\Gamma_\tot$ is depleted because of the reduction in the dominant width to $b \bar b$. These effects conspire to increase $\hR^g$, and at $\tan \beta=1$ this effect outweighs the decrease in $d_V$ for $\delta \lesssim \pi/4$ so that we see an increase for larger $\delta$.  Similar effects are seen in the top right panel, which shows the same decay but for vector boson fusion or vector boson associated production, but which is even less sensitive to $\tan \beta$. For the bottom panels, we show $R^g_b$ and $R^t_b$ with no singlet mixing, which illustrate that it is very hard to boost bottom production unless we go to very large $\tan \beta$ while simultaneously avoiding the decoupling limit.  Bottom production can be increased more effectively by going to large $\tan \beta$ in the case of gluon fusion as compared to top associated production because the gluon loop contains a bottom piece.  Thus, these panels are distinct as $\beta \to \pi/2$, but they are otherwise very similar.

\section{Loop Level Effects}\label{sec:loop}

With results for tree-level processes in hand, we can now consider the multi-faceted effects in the loop-mediated process $h \rightarrow \gamma \gamma$. The mixing angles $\delta$, $\beta$, and $\gamma$ and three loop parameters $\eta_{I,i}$ enter.  We will find again that the physics can be more easily extracted and understood by normalizing to the tree level rate to gauge bosons $R_V^j$.
The diphoton signal strength at leading order
is
\bea
d_\gamma & \simeq & \cA_{\gamma,V}  d_V + \cA_{\gamma, t} d_t + \cA_{\gamma,b} d_b + \underset{i\not \in {\rm SM}}{\sum} \cA_{\gamma,i} d_i, \label{dgamma2}
\eea
where we have approximated by only including the dominant SM loop contributions from the electroweak vector bosons and the top and bottom quarks.  Dividing both sides of \Eq{dgamma2} by $d_V$ and rearranging terms using the fact that $\cA_{\gamma, V} + \cA_{\gamma, t} + \cA_{\gamma,b} = 1$, we find that
\bea
\frac{d_\gamma}{d_V} & \simeq  & 1 + \cA_{\gamma, t} \left(\frac{d_t}{d_V}-1\right) + \cA_{\gamma,b} \left(\frac{d_b}{d_V}-1\right) + \frac{1}{d_V}\underset{i\not \in {\rm SM}}{\sum} \cA_{\gamma,i} d_i\\
& = & 1 + \epsilon(\beta) \tan \delta + \frac{1}{d_V}\underset{i\not \in {\rm SM}}{\sum} \cA_{\gamma,i} d_i, \label{AgammaoverAV}
\eea
where we have defined a function $\epsilon(\beta) = \cA_{\gamma, t} \cot\beta - \cA_{\gamma,b} \tan \beta $.  Simply squaring \Eq{AgammaoverAV}, we can recast the same information in terms of experimental observables, so
\begin{align}
\frac{R_\gamma^j }{R_V^j}
&\simeq  \Bigg|  1 + \ep(\beta) \tan \delta  + \frac{1}{d_V}\underset{i\not \in {\rm SM}}{\sum} \cA_{\gamma,i} d_i  \Bigg|^2. \label{rgamgen}
\end{align}The first, second and third terms in the above expression correspond to {\em (i)} the SM contribution,  {\em (ii)} the effect of mixing on the $t$ and $b$ Yukawas, and {\em (iii)} the effects of any additional charged particles beyond the SM.

\begin{figure}[t]
\centering
\includegraphics[width=0.6\textwidth]{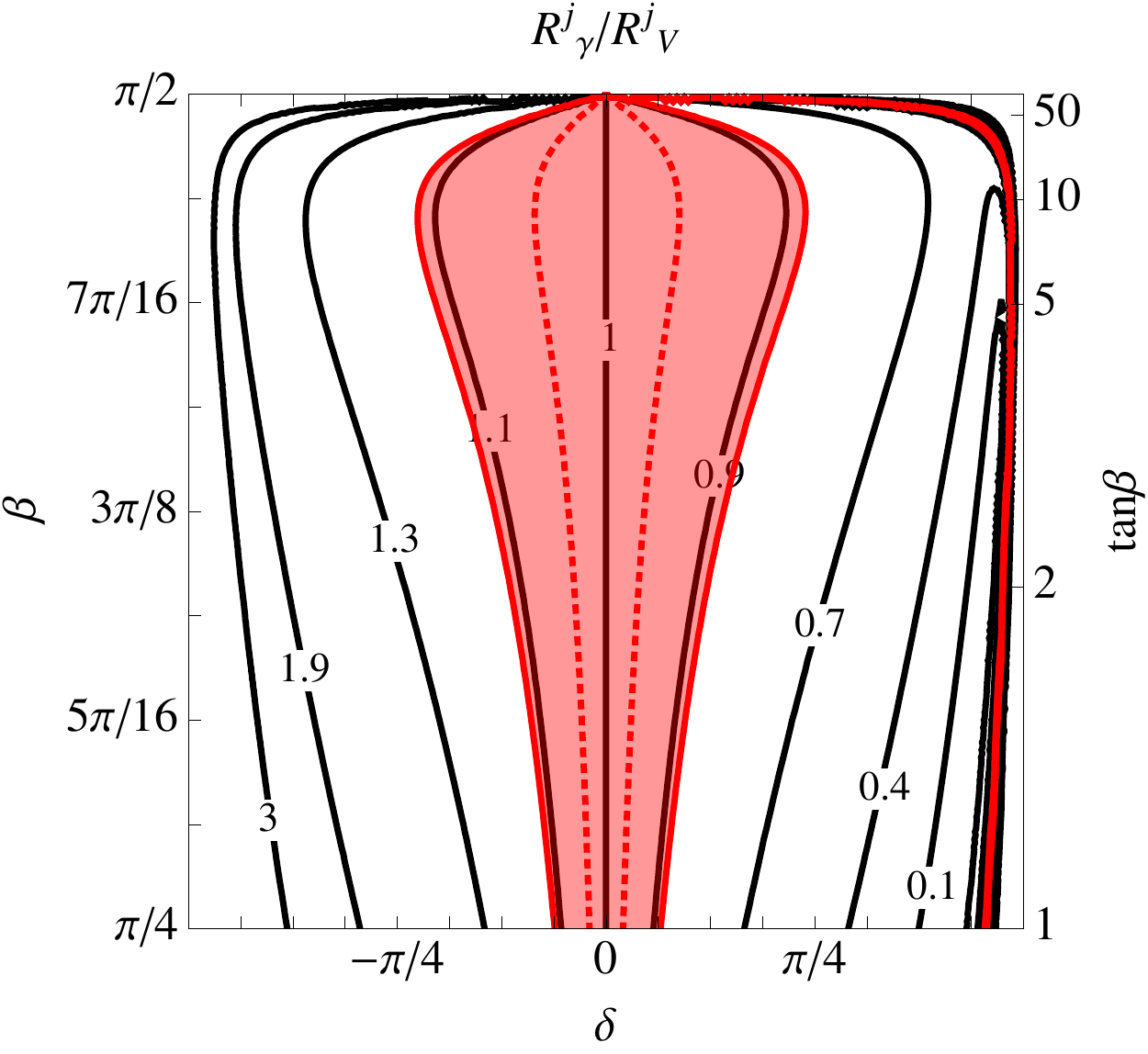}
\caption{ Contours of $R_\gamma^j/R_V^j$ in the plane of theoretical parameters $(\delta, \beta)$, with no new charged particles beyond the SM present. The red shaded regions are as in \Fig{tbtd}. }
\label{rgamV}
\end{figure}

Let us consider signal strengths in the case with no new charged particles beyond the SM, so that only mixing effects induce deviations from unity. Thus, only the contributions from $(i)$ and $(ii)$ are present in \Eq{rgamgen},
\beq
 \frac{R_\gamma^j}{R_V^j}  = \left|  1 +  \ep( \beta)  \tan \delta   \right|^2. \label{gamonV}
\eeq
In Fig.~\ref{rgamV} we map out contours of Eq.~\eqref{gamonV}. Comparing Figs.~\ref{bonVfig} and \ref{rgamV}, we see that $R_V^j$ is more tightly correlated with $R_\gamma^j$ than with $R_b^j$, since in the $(\delta,\beta)$ plane, $R^j_\gamma / R^j_V$ spans over a much more narrow range than $R^j_b / R^j_V$.  The reason for this correlation is obvious: $h\rightarrow \gamma\gamma$ is a process dominated by a $W$ loop, so it is highly correlated with the decay $h\rightarrow WW$.  Meanwhile, $h\rightarrow bb$ and $h \rightarrow WW$ are uncorrelated because the Higgs coupling to bottom quarks is controlled only by the $\phi_d$ component of $h$, while the Higgs coupling to electroweak vector bosons is controlled by both the $\phi_u$ and $\phi_d$ components of $h$.
Thus in order to decouple the Higgs rate to photons relative to the rate to gauge bosons, \emph{i.e.}~to push $R^j_\gamma / R^j_V$ far from unity, loop effects from new charged particles must be included. Breaking the correlation between these signal strengths will be one of the primary effects we are investigating, but loop effects will be critical for doing so.

\begin{figure}[t]
\begin{center}
\includegraphics[width=\textwidth]{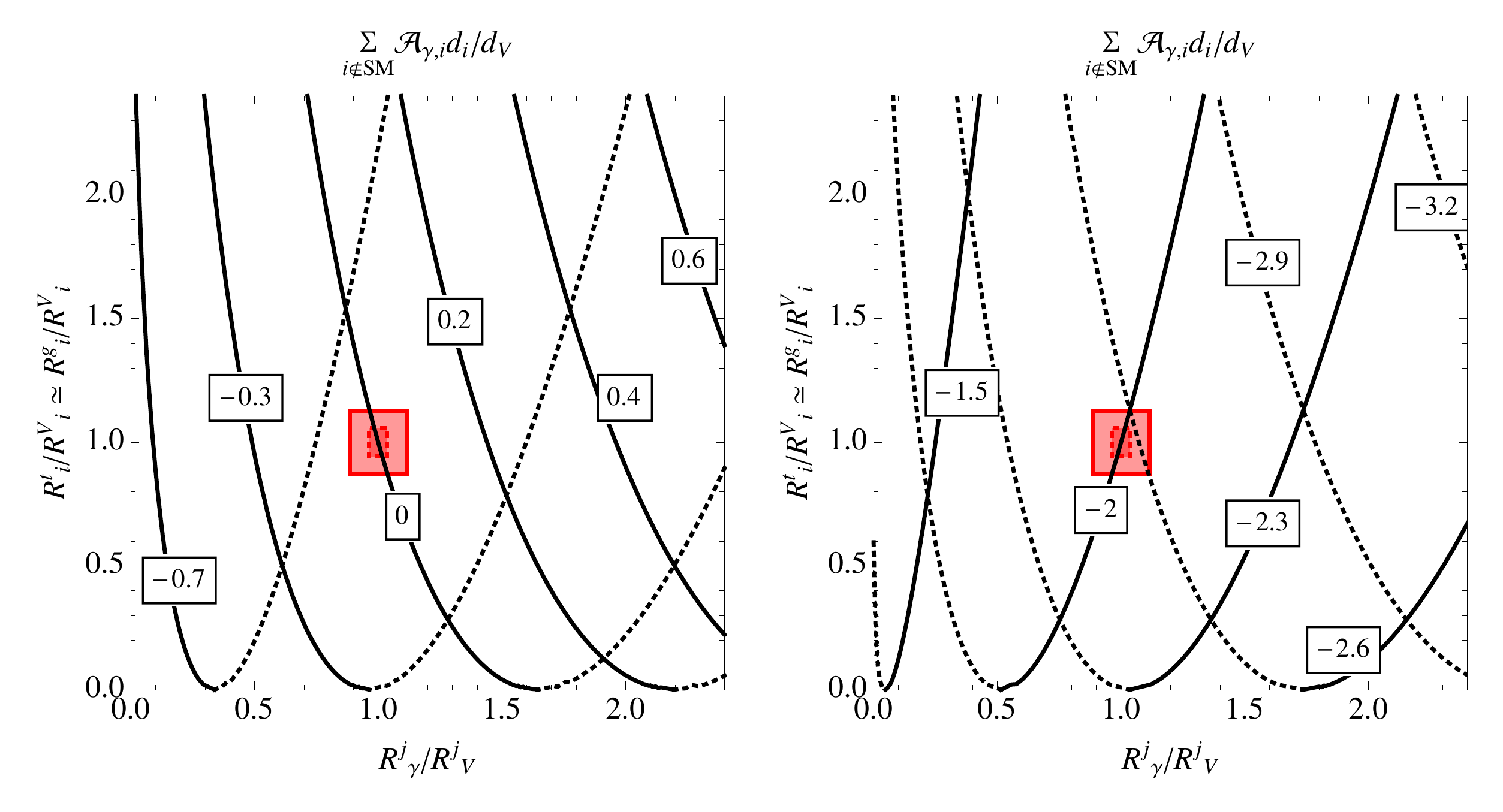}
\caption{Taking the low $\tan \beta$ limit, we plot the new physics contribution to the diphoton rate as a function of $R_\gamma^j/ R_V^j$ and $R^g_i/ R^V_i.$ The left (right) panel shows values of $\sum_{\rm i\not \in {\rm SM}} {\cal A}_{\gamma,i} d_i/d_V$ 
for which $d_\gamma/d_V$ is positive (negative).  In both panels, solid (dotted) lines show values of $\sum_{\rm i\not \in {\rm SM}} {\cal A}_{\gamma,i} d_i/d_V$ for which $d_g/ d_V$ is positive (negative).  The red shaded regions are as in \Fig{tbtd}. We use the uncertainty in $R^g_i/R^V_i$ rather than $R^t_i/R^V_i$ since the former will be much better measured, as shown in \Tab{tabunc}.}
\label{ccx}
\end{center}
\end{figure}

Let us now investigate these loop effects. As expected, in the decoupling limit, $\delta \rightarrow 0$, the effects of {\em (ii)} vanish but {\em (iii)} can still play an important role. We plot the unknown quantity $\sum_{\rm i\not \in {\rm SM}} {\cal A}_{\gamma,i} d_i/d_V$ in \Fig{ccx} as a function of the ratios of signal strengths that appear in \Eq{rgamgen}, taking the low $\tan \beta$ limit such that $d_g \simeq d_t$, and thus $\ep(\beta)\simeq \cA_{\gamma,t} (d_g/d_V -1)$.  Because the signal strengths are related to $d_\gamma$ and $d_g$ by squaring, the latter are only fixed up to a sign ambiguity.
In the left (right) panel we show contours of $\sum_{\rm i\not \in {\rm SM}} {\cal A}_{\gamma,i} d_i/d_V$ where the sign of $d_\gamma/d_V$ is positive (negative).   In both panels, the solid (dotted) curves show regions where $d_g/ d_V$ is positive (negative).  Note that negative values of $d_g/d_V$ or $d_\gamma/d_V$ require large effects from new physics, and are far from SM-like.

Let us now discuss in more detail how {\em (iii)} acts on the $h\rightarrow \gamma \gamma$ rate.
As shown in \Eq{Anumerical}, the $W$ loop contribution to the Higgs coupling to photons destructively interferes with but dominates over the top and bottom quark loop contributions.  Regardless of whether the new particles $i$ are scalars or fermions, they have the greatest effect if their loop contributions constructively interfere with the $W$ loop.  Thus, large negative values of the new $d_i$ will lead to the largest enhancements.  The spin of a new particle $i$ does play an important role in the Higgs coupling to photons, however, because the quantities $\cA_{\gamma, i}$ asymptote towards beta function coefficients in the large $m_i$ limit. A Dirac fermion $\psi$ contributes more strongly to running than a complex scalar $\phi$ of the same quantum numbers, since $b_{1/2}=4b_0$, and we find that for large masses
\bea
4\eta_{I,\phi}= \eta_{I,\psi},
\eea
which signifies that for a fermion and a scalar with equal masses, the scalar must couple $4$ times more strongly to the Higgs in order to account for the same effect on $h\rightarrow \gamma\gamma$ as the fermion.  Equivalently, for a fermion and a scalar with the same coupling to the Higgs boson, the scalar must be one quarter as massive in order to accommodate the same effect on $h\rightarrow \gamma\gamma$ as the fermion.

We will henceforth assume the existence of a single, unit charged fermion  $\psi$ with a mass that satisfies $2m_\psi \gg m_h$, so that its corresponding loop coefficient $\cA_{\gamma,\psi}$ is fixed by the beta function coefficient of the new particle, $\cA_{\gamma,\psi} \to b_{1/2}=4/3$.    We treat $\eta_{I,\psi}$ as a free parameter which can vary independently of $m_\psi$.  As noted earlier, only keeping renormalizable terms in the action typically suggests a range $|{\eta}_{I,\psi}| \lesssim 1$ if the new particle is a fermion.
\begin{figure}[t]
\begin{center}
\includegraphics[width=.5\textwidth]{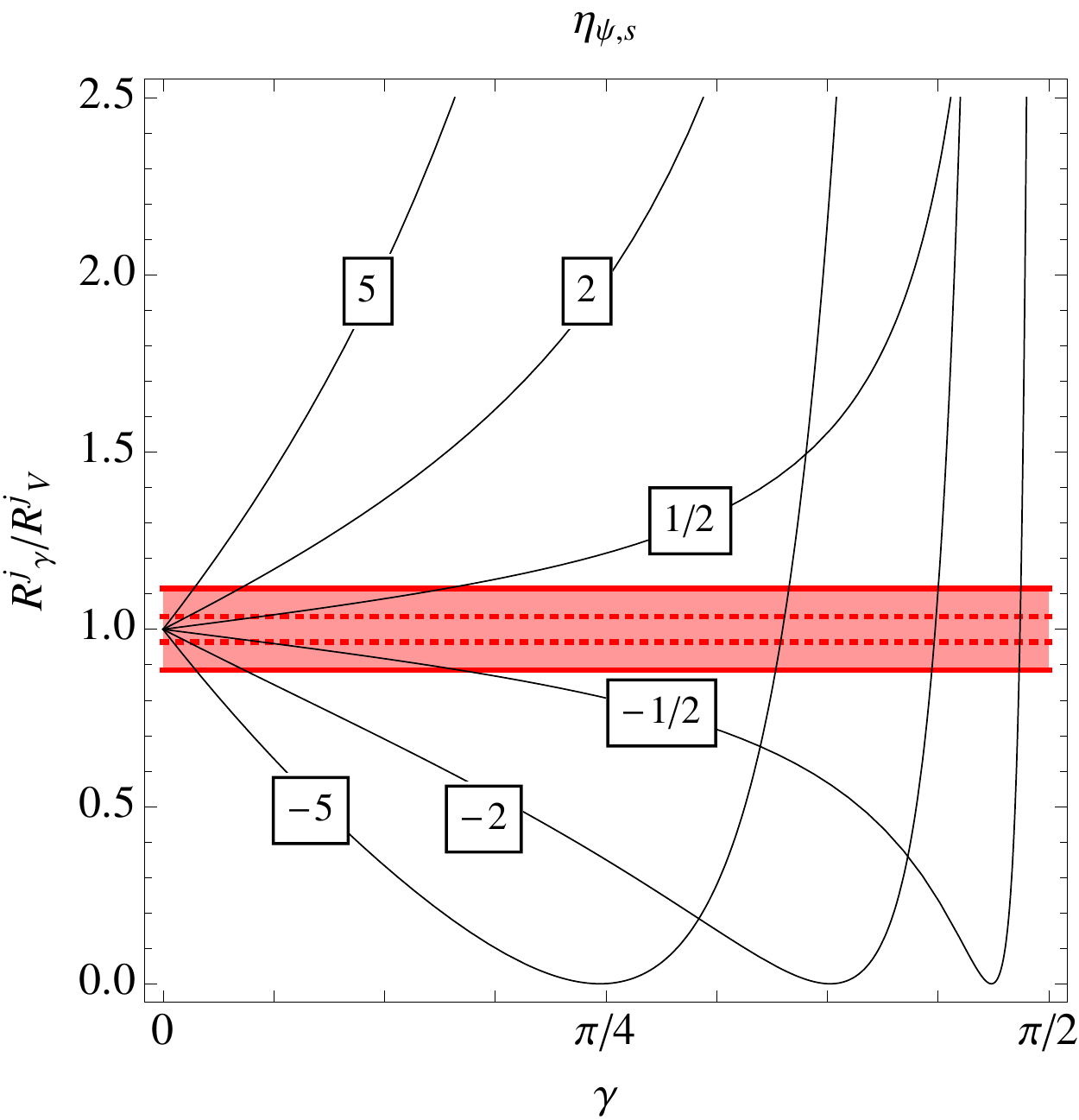}
\caption{Contours of $\eta_{s,\psi}$ as a function of  $(R_\gamma^j/R_V^j,\gamma)$, in the decoupling limit $\delta=0$. The red shaded regions are as in \Fig{tbtd}.}
\label{etabetadec}
\end{center}
\end{figure}
The ratio of $R_\gamma^j/R_V^j$, including the effects of mixing and a single unit-charged fermion is approximately
\begin{multline} \label{gamonVws}
 \frac{R_\gamma^j}{R_V^j} \simeq \Bigg|  1   -  \pL 0.282 \cot \beta - 6.14 \times 10^{-3} \tan \beta  \pR  \tan \delta \\
-  0.204 \frac{ \sin( \beta+ \delta) \eta_{u,\psi} + \cos( \beta+ \delta)  \eta_{d,\psi} -  \tan \gamma ~\eta_{s,\psi} }{\cos \delta }   \Bigg|^2 . 
\end{multline}
To get a sense for the effect of charged particles we consider the decoupling limit, $\delta \to 0$, and we assume that the coupling goes only through the singlet, so that \Eq{gamonVws} vastly simplifies to  ${R_\gamma^j} / {R_V^j} \simeq |1 + 0.204\tan \gamma \, \eta_{s,\psi}|^2$, which is depicted in Fig.~\ref{etabetadec}.  We see that $R_\gamma^j/R_V^j $ is more sensitive to the values of the mixing angles for larger magnitudes of $\eta_{s,\psi}$, which nicely illustrates the interplay between the tree-level mixing and loop-level coupling effects: we see here that ramping up the tree-level effect of the mixing angle is ineffective unless the coupling of the loop particle is also appreciable. Without this combined effect, the mixing angles must be very large in order to produce sizable effects on Higgs properties.

\begin{figure}[t]
\centering
\includegraphics[width=.6 \textwidth]{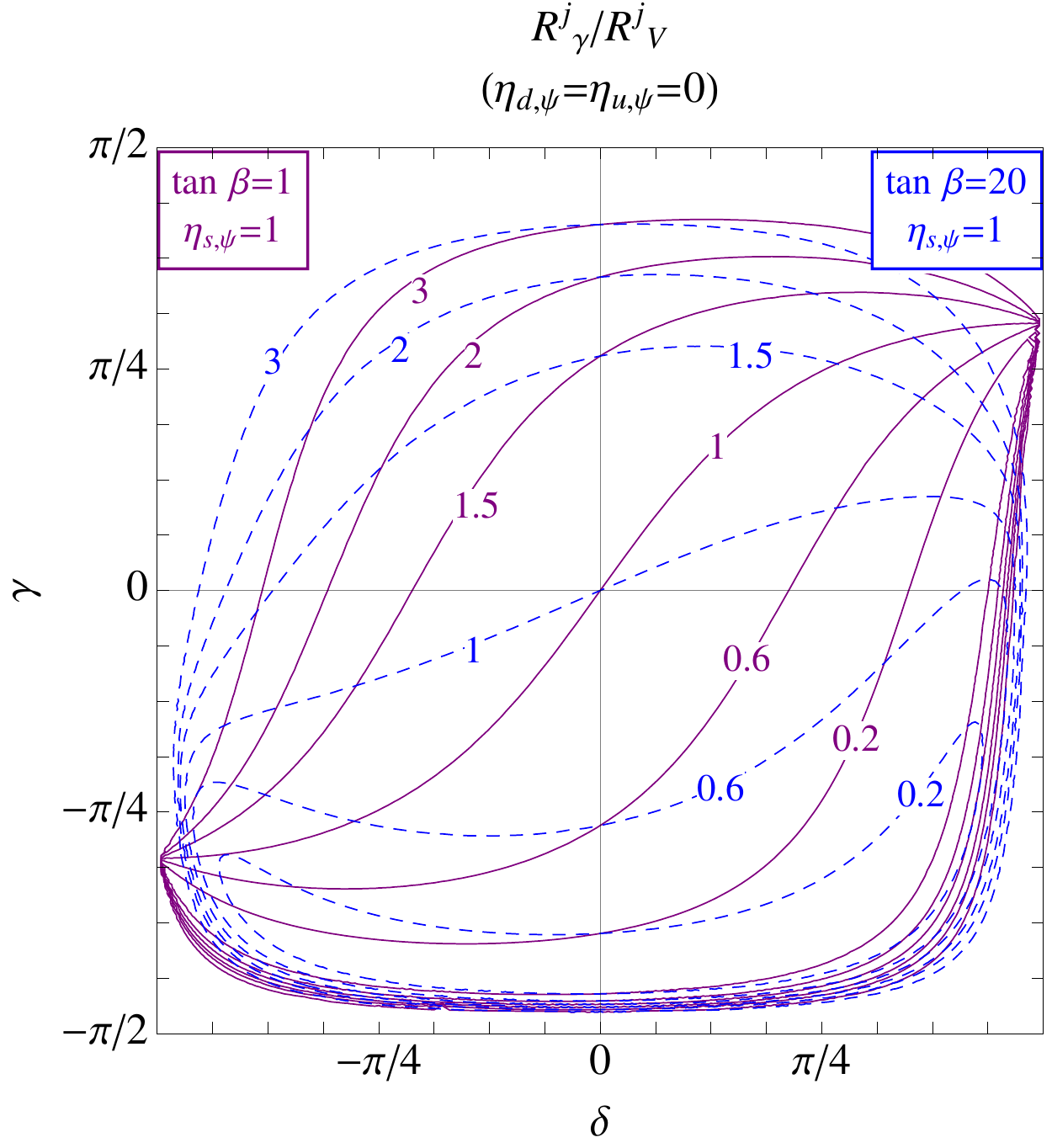}
\caption{Contours of $R_\gamma^j/R_V^j$ as a function of $(\delta, \gamma)$ for fixed values of $\tan \beta$ and $\eta_{s,\psi}$ for a unit-charged fermion, and $\eta_{d,\psi}=\eta_{u,\psi}=0$. }
\label{deltagam}
\end{figure}

\begin{figure}[t]
\centering
\includegraphics[width=\textwidth]{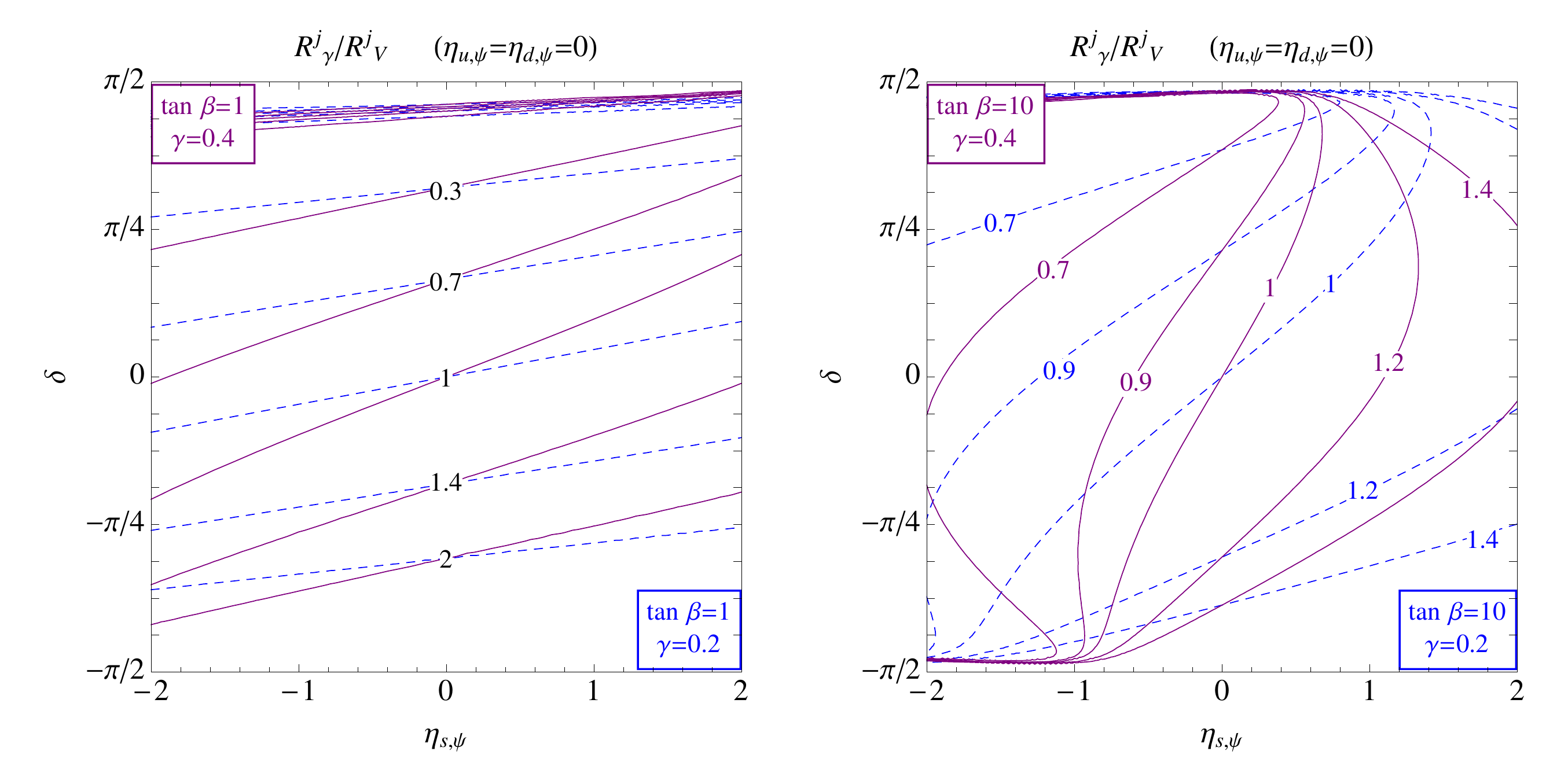}
\caption{Contours of constant $R_\gamma^j/R_V^j$ for fixed values of $\tan \beta$ using a single fermion loop particle that couples only to the singlet component of the Higgs. We take $\gamma = 0.2~ (0.4)$, which represents the 68\% (95\%) confidence level limits on the amount of the singlet in the lightest physical Higgs. The pinching behavior should occur in the left panel as well, but requires $|\eta_{s,\psi}| > 2$. }
\label{etadel}
\end{figure}

We now examine in greater depth the interplay of the mixing angles with the loop effects in Figs.~\ref{deltagam}--\ref{etadel}, which depict contours of $R^j_\gamma / R^j_V$, as shown in Eq.~\eqref{gamonVws}. The features in these plots are determined by the functional relationships in Eq.~\eqref{gamonVws}, 
but we see a few broad patterns.  As noted before,  $\cO(0.3) \lesssim R_\gamma^j/R_V^j \lesssim \cO(3)$ obtains in the vast majority of parameter space. This is expected because the $W$ boson loop dominates the Higgs coupling to photons, so that $R_\gamma^j$ is constrained to remain within a factor of a few of $R_V^j$ unless some cancellations occur. To achieve the largest possible separation of these rates we require one of two possibilities.
\begin{itemize}
\item The Higgs couples substantially to new charged particles.   The ratio $R^j_\gamma / R^j_V$ is enhanced when the new particle interferes constructively with the $W$ loop contribution, which occurs for the largest positive values of the product $\tan \gamma \, \eta_{s,\psi}$ or for negative (positive) values of $ \eta_{u,\psi}$ and $\eta_{d,\psi}$ if $\delta < \pi/2 -\beta~({\rm if~}\delta > \pi/2 -\beta)$.
\item The Higgs mixes with scalars which substantially modify its coupling to top and bottom quarks.   The ratio $R^j_\gamma / R^j_V$ is enhanced (suppressed) if $\delta < 0$ ($\delta >0$).  As $\tan\beta$ increases, these effects diminish until the bottom contribution becomes important and the effects strengthen again.  
\end{itemize}
These combined effects allow a non-negligible separation of the $\gamma$ and $W$ signal strengths without resorting to extreme values of the couplings or mixing angles.

These two simple phenomena describe most of the broad features we see in Figs.~\ref{deltagam}--\ref{etadel}.  However, a few other important features can occur in small slices of parameter space where certain Higgs rates go to zero: for example, due to vanishing Higgs couplings to photons (at special values of the mixing angles or the couplings), to the massive electroweak vector bosons (at $\delta\to \pm \pi/2$), or to all SM particles (at $\gamma \to \pm \pi/2$).  These limits cause $R_\gamma^j/R_V^j$ to either diverge or vanish, thus deviating greatly from the general correlation $R_\gamma^j/R_V^j \sim \cO(1)$.  
We begin with Fig.~\ref{deltagam}, where we see the dependence of $R^j_\gamma /R^j_V$ as a function of $(\delta, \gamma)$ given a single, unit-charged fermion that couples exclusively to $\phi_s$ with strength $\eta_{s,\psi} =1$.  The Higgs coupling to SM particles will vanish as $\gamma \simeq \pm \pi/2$, but the Higgs couplings to photons will still be mediated in this case by particles that couple to the $\phi_s$.  Thus as long as $\eta_{s,\psi} \neq 0$, then as we approach $\gamma \simeq \pm \pi/2$, $R_V^j$ will decrease faster than $R^j_\gamma$, and the ratio in Eq.~\eqref{gamonVws} diverges. As we approach $\gamma \simeq - \pi/2$, however, there is a special intermediate value of $\tan \gamma$ below which all the contours are tightly packed, which occurs because the Higgs coupling to photons vanishes but the coupling to the massive electroweak vector bosons does not. The photon coupling can only vanish if the new charged particle causes total destructive interference in the photon loop.
If $\eta_{s,\psi} < 0$ we would find the clustering at positive values $\tan \gamma$ where the singlet contribution could again interfere destructively with the $W$ loop.  In Fig.~\ref{deltagam}, there is also a ``pinching'' feature at the corners of the plot near the anti-coupling limit, $\delta\simeq \pm \pi/2$.  These features arise because at those points the Higgs couples neither to photons nor to vectors.

We see very similar behavior in Fig.~\ref{etadel}, where we again consider a single unit-charged fermion which couples to the $\phi_s$ component of the Higgs. The clustering and pinching behaviors as $\delta \to \pm \pi/2$ are precisely analogous to those described in Fig.~\ref{deltagam}. The behavior of $R_\gamma^j/R_V^j$ is in fact symmetric in $\eta_{s,\psi}$ and $\tan \gamma$ because this ratio only depends on the product $\tan \gamma \, \eta_{s,\psi}$--the plots are visibly different only because we have plotted $\gamma$ as the axis in \Fig{deltagam} rather than $\tan \gamma$.   

We also have chosen discrete values of $\gamma=0.2,0.4$. As follows from the general discussion and as shown in \Fig{etabetadec}, we see that for larger mixing angles $\gamma$ the diphoton rate is more sensitive to $\eta_{s,\psi}$ because the new charged particle can now be more strongly coupled to the physical Higgs. In the same way, the $\delta$ dependence is sensitive to the value of $\beta$. Because the top quark loop destructively interferes with the $W$ loop we see an enhanced diphoton rate where the $t$ interference is most suppressed -- this occurs with the smallest $\tan \beta$ and when $\delta<0$. This effect is diminished at larger $\tan \beta$ so that the contours are less steeply inclined against the $\delta$ direction and more inclined against $\eta_{s,\psi}$, as is seen by comparing the left and right panels of Fig.~\ref{etadel}.

\section{NMSSM }\label{sec:NMSSM}

The NMSSM is one of the best studied and most well-motivated examples of a singlet extension to the standard 2HDMs.  Here we show how our formalism facilitates the analytic extraction of non-SM-like Higgs production and decay. We use the standard MSSM superpotential (retaining the $\mu$ term) enhanced by singlet contribution
\beq
W\supset \pL \mu+ \lambda S\pR H_u \cdot H_d + \frac12 \mu_s S^2 + \frac13 \kappa S^3.
\eeq
We also use standard terminology for the soft SUSY-breaking terms, such that the soft SUSY-breaking Lagrangian contains soft Higgs terms 
\begin{multline}
\cL_{\rm soft}\supset m_{H_u}^2|{H_u}|^2+m_{H_d}^2|{H_d}|^2+m_S^2 |S|^2  + \pL  B \mu H_u \cdot H_d + \half B_s \mu_s S^2  + {\rm h.c.} \pR \\+ \sqrt{2} \pL  A_\lambda  S H_u \cdot H_d  - \frac13 A_\kappa S^3 + {\rm h.c.} \pR.
\end{multline}
Mapping the phenomenological considerations of \Secm{Framework}{sec:loop} onto the NMSSM parameter space here, the only candidate particles within the NMSSM which could in principle be important for our analysis are: squarks, most importantly the $\widetilde{t}$; sleptons, most importantly the $\stau$; the charged Higgs, which has very tight direct and indirect search constraints that require its mass to be $\gtrsim \cO(300\gev)$, but which does not automatically decouple in the high mass limit; and the chargino, $\chi^\pm$. The $\widetilde{t}$'s are colored, which generically induce large differences in the gluon fusion and vector boson fusion production channels which are not observed \cite{fit}. More importantly, both the stops and staus require very low masses and very large mass mixings to enhance the Higgs diphoton rate, which can destabilize the vacuum by inducing charge- or color-breaking minima \cite{Reece:2012gi,staus,staustab}.   In general, vacuum stability can also arise from purely electroweak charged or singlet states if one requires very large modifications to Higgs properties \cite{nima}, or the requirement of thermal dark matter \cite{Cheung:2012nb}.  For more modest deviations from a SM-like Higgs, we can ignore more general questions of vacuum stability in the NMSSM \cite{Agashe:2012zq}  and study the case of the chargino and the charged Higgs.

\subsection{Masses And Kinematics}

Next, we focus on changes to Higgs production and decay in the NMSSM.  The one loop corrections to $h\rightarrow \gamma \gamma$ will come from the charged Higgs $H^\pm$ or the charginos $\chi^\pm$.  A multiplet of new physics states of equal charge labeled by $i$ contribute to the Higgs coupling to photons with a weight
\bea
\eta_I &=& \sum_i \eta_{I,i} = \sum_i \frac{v}{m_i} \frac{\partial m_i}{\partial v_I} = v\frac{\partial }{\partial v_I} \log \det \cM ,
\eea
following \Eq{eta}.  Here $\cal M$ is the mass matrix for the particles $i$, so the eigenvalues of $\cal M^\dagger \cM$ are $m_i^2$.

The charged Higgs mass is 
\beq \label{mhch}
m_{H^\pm}^2= \bL \pL  A_\lambda  + \frac{ \lambda \mu_s}{ \sqrt{2}} + \frac{\lambda \kappa v_s}2 \pR v_s +  B \mu \bR \frac{v^2}{v_u v_d} + \frac{v^2}2 \pL \frac{g^2}2 - \lambda^2 \pR ,
\eeq
and the chargino mass mixing matrix is
\beq \label{mfmat}
\cM_{\chi^\pm} = \pL \begin{array}{cc} M_2 &  g v_u/\sqrt{2} \\  g v_d/\sqrt{2} & \mu_{\rm eff}  \end{array} \pR,
\eeq
where $M_2$ is the wino mass parameter and $\mu_{\rm eff}=\mu + \lambda v_s/\sqrt{2}$. The determinant of this matrix is the product of the two chargino masses, which we will call $\overline{m}^2$:
\beq \label{mcino}
\pL \overline{m}^2 \pR^2 = \pL m_{\chi^\pm_1} m_{\chi^\pm_2} \pR^2 = \det \pL \cM_{\chi^\pm}^\dag \cM_{\chi^\pm} \pR   = \pL M_2 \mu_{\rm eff} -  m_W^2 \sin 2\beta \pR^2.
\eeq
We will use $\overline{m}^2$ to parameterize the chargino loop effects. Current chargino constraints from LEP simply require $m_{\chi^\pm_1} \gtrsim 103.5 \gev$ for generic neutralino masses or $m_{\chi^\pm_1} \gtrsim 92 \gev$ for nearly degenerate chargino and neutralino masses \cite{LEP}.

\begin{figure}[t]
\begin{center}
\includegraphics[width=\textwidth]{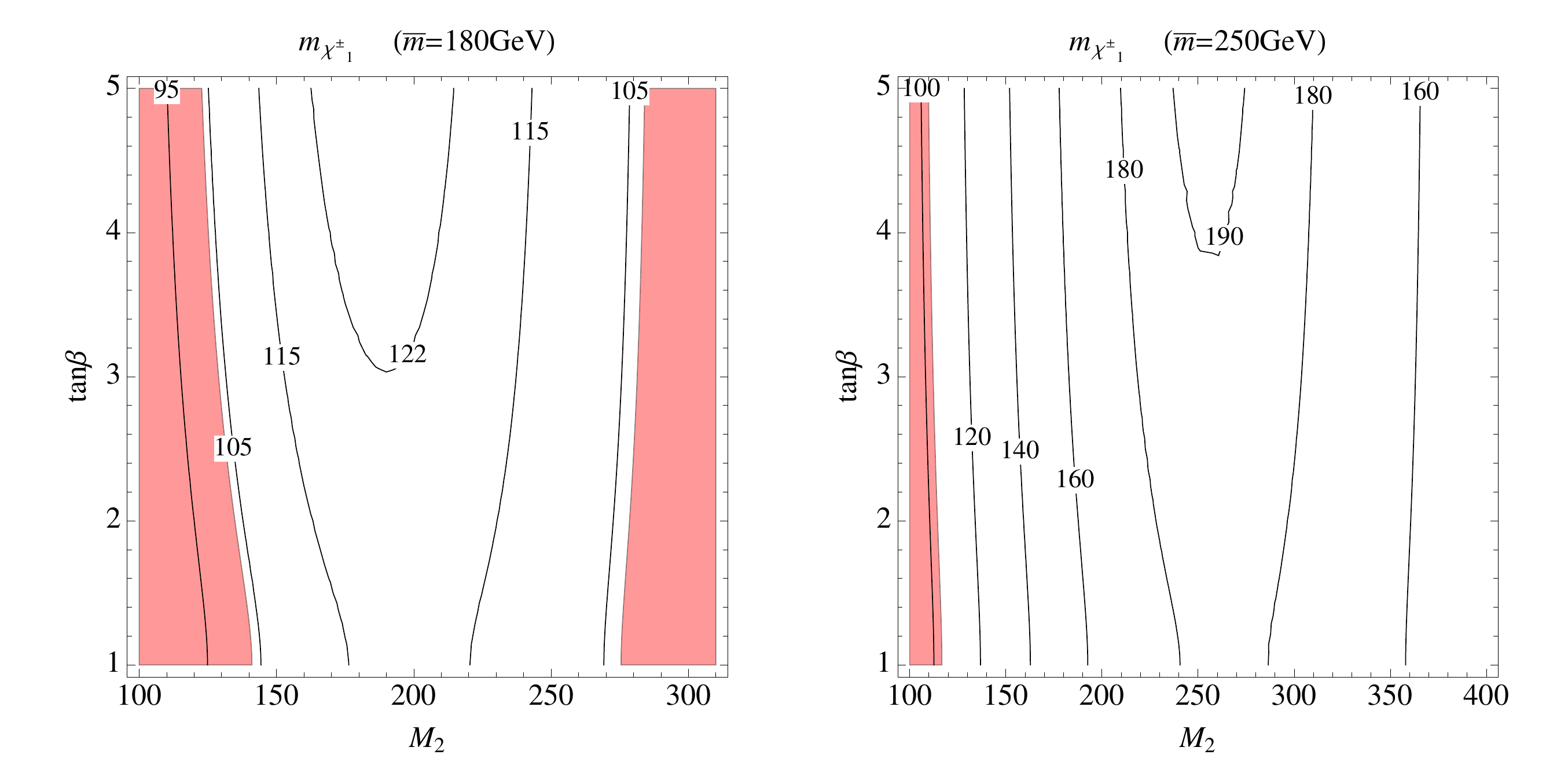}
\caption{Contours of constant $m_{\cino_1}$ for fixed $\overline{m}$. Regions that fail LEP bounds are shaded red.}
\label{cinomass}
\end{center}
\end{figure}

In Fig.~\ref{cinomass} we show contours of constant $m_{\chi^\pm_1}$ for realistic fixed values of $\overline{m}$. For $M_2 \lesssim \overline{m}$, the lightest chargino is predominantly $\wino$ so that the $\cino_1$ mass grows linearly with $M_2$, while the opposite is true for large $M_2$ where the lightest chargino is mostly $\hino$. The transition from $\wino$ to $\hino$ occurs around $\overline{m}$, and at this point the off-diagonal terms in $\cM_{\cino}$ become important so that there is increased sensitivity to $\tan \beta$ in this region. Away from this feature in the parameter space the chargino masses are well split so that $m_{\cino_{1,2}} \sim \mu_{\rm eff}/\sqrt{2}, M_2$ and there is less dependence on $\tan \beta$.

\subsection{Signal Strengths}

\begin{figure}[t]
\begin{center}
\includegraphics[width=\textwidth]{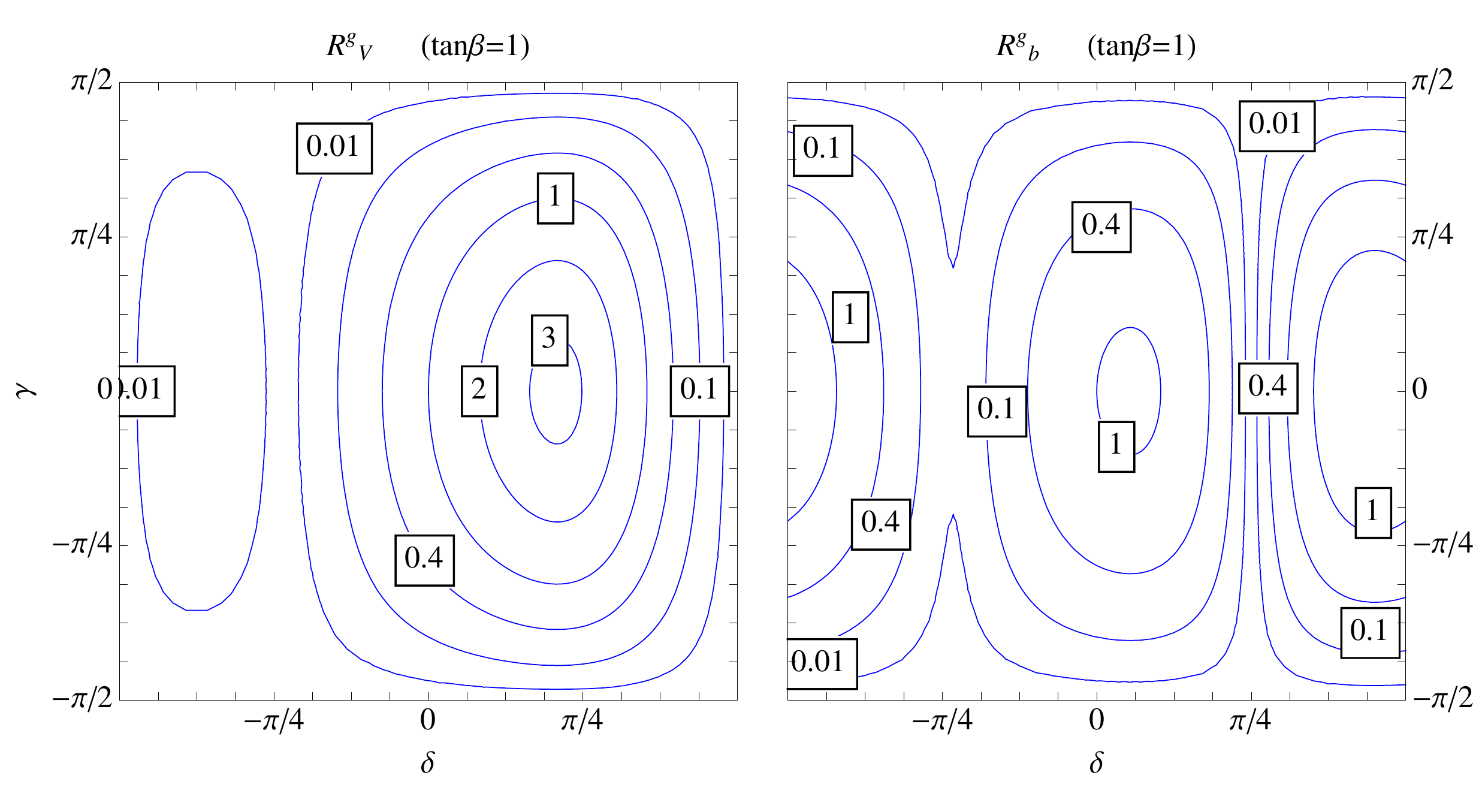}
\caption{Contours of ({left}) $R_V^g$ and ({right}) $R_b^g$ for $\tan \beta=1$.}
\label{rSMnorm}
\end{center}
\end{figure}

In Fig.~\ref{rSMnorm} we show contours of $R_{V,b}^g$ in the $\delta-\gamma$ plane in order to show which values of these parameters give rise to reasonable signal strengths. We take a fixed value of $\tan \beta=1$ since, as shown in \Figs{bonVfig}{rjv}, the $\beta$ dependence is mild at low $\tan \beta$. Near the origin---which represents the no-mixing, exact decoupling limit---we see that both of the $V$ and $b$ signal strength modifiers are highly sensitive to the departure from the decoupling limit, but not as sensitive to the amount of mixing. This arises because, as noted above, the mixing dependence in $\hR^j$ effectively decouples for small values of the singlet mixing angle since both $\sigma(jj\rightarrow h)$ and $\Gamma_{{\rm tot}}$ are proportional to $\cos^2 \gamma$. The only $\gamma$ dependence in $R_{V,b}^j$ therefore comes from the actual particle couplings, $d_{V,b}$. In contrast, both $\hR^j$ and the couplings carry strong sensitivity to $\delta$. Since we observe Higgs decays in rough overall agreement with the SM expectation, we assume that the mixing angles are near the decoupling limit. For simplicity we will take $\delta=0$ in the plots below to guarantee general agreement with the SM values, but we will allow $\gamma$ to vary in order to illustrate the effects on new particles that couple through the singlet.

The Higgs coupling vectors of Eq.~\eqref{coupling} for the chargino and the charged Higgs are
\begin{align}
{\eta}_{I ,\chi^\pm}&=  \frac{2m_W^2}{\overline{m}^2}  \pL -\cos \beta,- \sin \beta, \frac{\lambda M_2}{\sqrt{2} g m_W} \pR, \label{gencino}
\\  \label{genhpm} {\eta}_{I,H^\pm}&= \Bigg( \frac{m_W^2}{m_{H^\pm}^2} \frac{1-2\lambda^2/g^2}{2 \sin \beta}  - \frac{\cos 2\beta}{2\sin \beta},  \frac{m_W^2}{m_{H^\pm}^2} \frac{1-2\lambda^2/g^2}{2 \cos \beta} + \frac{\cos 2\beta}{2\cos \beta},\\ &\qquad \qquad \frac{2 m_W}{g m_{H^\pm}^2} \frac{A_\lambda + \lambda \pL \kappa v_s + \mu_s/\sqrt{2}\pR}{\sin 2\beta} \Bigg) \nonumber.
\end{align}
The constant terms in the $H^\pm$ coupling vector ensure that the charged Higgs does not automatically decouple even as $m_{H^\pm} \to \infty$. However, for very large mass, as required by direct and indirect searches, we can make additional statements about the strength of this coupling. We find that $d_{H^\pm}$ is proportional to the mixing angle $\delta$ (which tends to be small when $m_{H^\pm}$ is large) and to $\cot 2 \beta$ (which goes to 0 as $\tan \beta \to 1$). Combining these scaling arguments with the relative size of the fermion and scalar beta functions ($b_{1/2}=4b_0$), the $H^\pm$ is ineffective compared to the $\cino$ throughout the bulk of the parameter space we are interested in.

The ratio of the diphoton and vector signal strength modifiers in the NMSSM as a function of the $\chi^\pm$ couplings is 
\begin{multline}
\pL \frac{R_\gamma^j}{R_V^j} \pR_{\rm NMSSM} \simeq  \Bigg| 1 + \epsilon(\beta) \tan\delta  -   \cA_{\gamma,\chi^\pm} \left( \frac{2 m_W^2}{\overline{m}^2} \frac{\sin(2\beta + \delta)}{\cos \delta}  + \frac{\sqrt{2} \lambda m_W M_2}{ g~ \overline{m}^2 }  \frac{ \tan \gamma}{ \cos \delta} \right) \Bigg|^2.
\end{multline}
From this expression, we see that the photon and vector rates can be different from each other most effectively for large positive $\gamma$, large $\delta$, and small $\tan \beta$. The $\ep(\beta)$ piece vanishes in the decoupling limit and the singlet piece does not contribute if there is no mixing, but even in these combined limits the ratio still delivers an increase over the SM prediction by a factor $\sim 4m_W^2/5\overline{m}^2 \sim \cO (20\%)$ because of the presence of the additional loop particle. As anticipated above, the couplings $\eta_{I,\cino}$ are not independent of the chargino mass parameters. Interestingly, the sign of this effect works to give an enhancement of $R_\gamma^j$ as compared to $R_V^j$.

\begin{figure}
\begin{center}
\includegraphics[width=\textwidth]{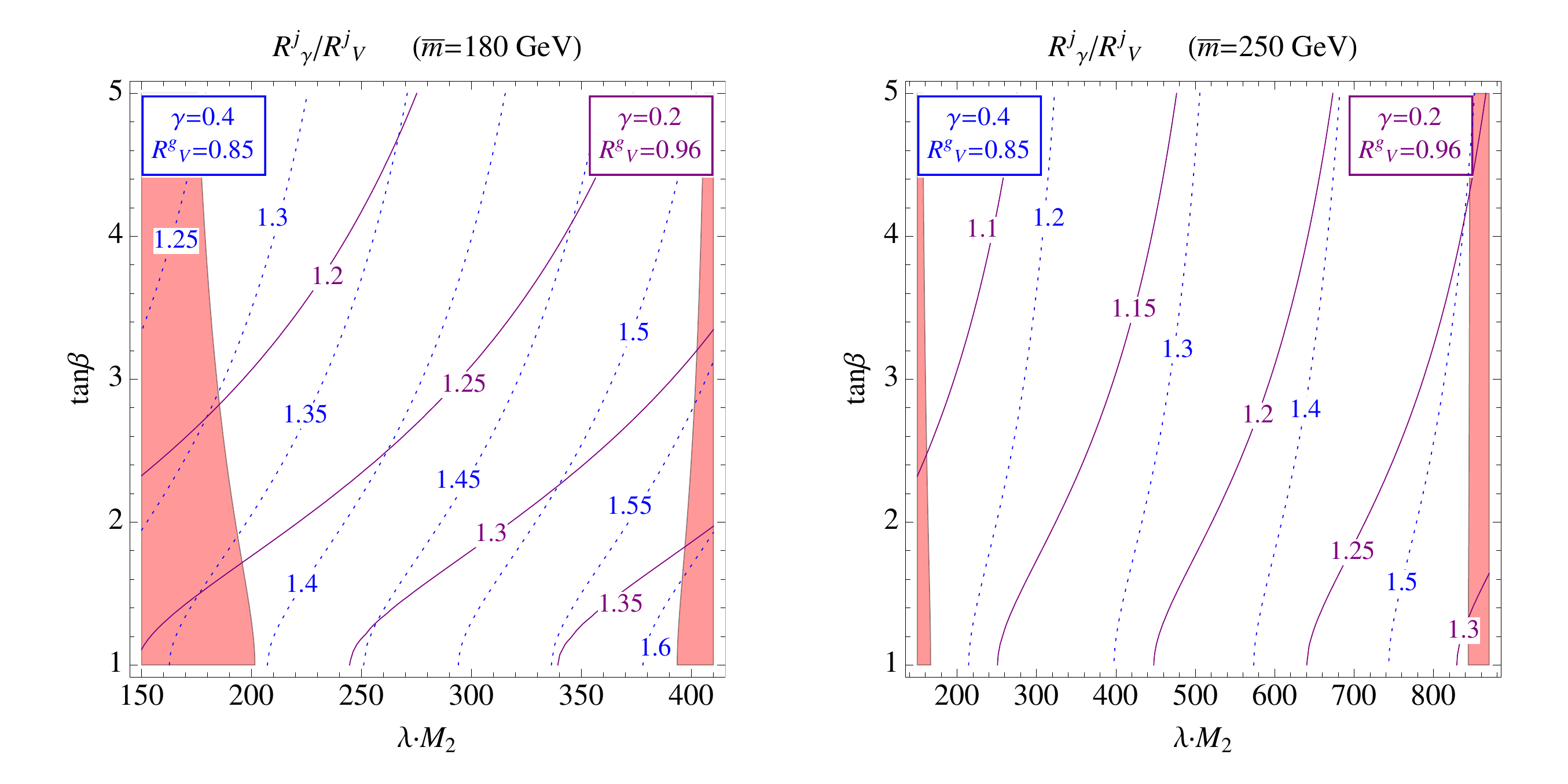}
\caption{Contribution from the $\chi^\pm$ loop in the decoupling limit with $\gamma=0.2,0.4$ for ({left}) $\overline{m}=150\gev$ and ({right}) $\overline{m}=250\gev$. In the shaded red regions, $m_{\cino_1} \leq 103.5\gev$ for $\lambda=0.7$. }
\label{cino}
\end{center}
\end{figure}

\begin{figure}[h]
\begin{center}
\includegraphics[width=\textwidth]{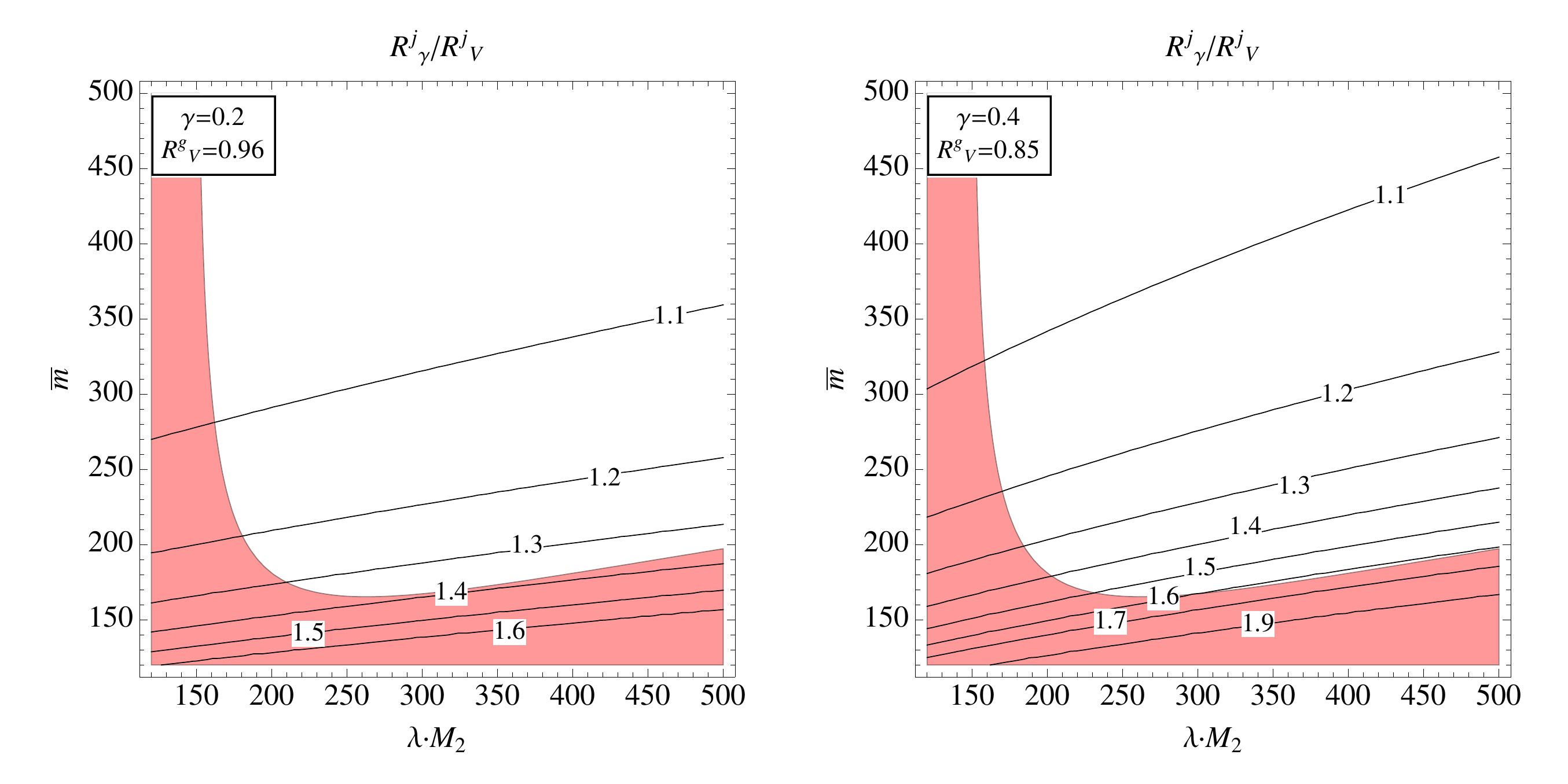}
\caption{Same as Fig.~\ref{cino} but in the $\lambda M_2 - \overline{m}$ plane and with $\tan \beta=1$.}
\label{cino2}
\end{center}
\end{figure}

In Fig.~\ref{cino} we display the values of $R_\gamma^j/R_V^j$ resulting from $\chi^\pm$ loop effects as a function of $\tan \beta$ and $\lambda M_2$, which respectively parameterize the size of the couplings to the doublet and singlet Higgses: $\eta_{u,\chi^\pm } \sim \eta_{d,\chi^\pm } \sim \tan \beta$ and $\lambda M_2 \sim \eta_{s,\chi^\pm}$. In Fig.~\ref{cino2} we show this ratio in the $\lambda M_2 -\overline{m}$ plane fixing $\tan \beta$ to 1. In both figures we shade the region for which $m_{\cino_1} \leq 103.5\gev$ when $\lambda=0.7$. Across all panels, we see that low $\overline{m}$, low $\tan \beta$ and large $\lambda M_2$ drive an enhanced $R_\gamma^j$ over $R_V^j$.  Within each panel, we see that the rate becomes more sensitive to the singlet coupling as we increase $\gamma$. We also see that $ M_2 \gtrsim \overline{m} $ is preferred, so that $\hino$-like charginos are more effective at driving this rate difference. The $\tan \beta$ dependence is also simple to understand, since in the decoupling limit it only enters through the $\sin 2\beta$ dependence of the doublet Higgs coupling, which is maximized at minimal $\tan \beta$. For instance, in the left panel of Fig.~\ref{cino} (where $\overline{m}=180\gev$) we see $R_\gamma^j/R_V^j \sim 1.6$ around $\lambda M_2 =380 \gev$ and $\tan \beta=1$. From \Fig{cino2} we see that $\lambda M_ 2 \geq \overline{m}$ leads to the greatest enhancements in $R_\gamma^j/R_V^j$, rising up to $R_\gamma^j/R_V^j \simeq 1.6$ for the example point picked above.

\section{Conclusions}

Higgs boson physics will be a critical component of the ongoing experimental effort at the LHC.  While a definitive picture of this newly discovered particle has yet to emerge, experimental errors will shrink, and could offer hints of new dynamics at the electroweak scale in years to come.
The focus of this paper has been on theories which have a greater likelihood of being discovered indirectly through modified Higgs properties, as opposed to direct LHC searches.
We have investigated a simple parameterization of new physics effects from new electroweak charged and singlet particles which couple to or mix with the Higgs boson.  Our formalism has the advantage that it can encompass a number of different electroweak extended models with relatively few parameters.  Furthermore, observables such as strength modifiers for various Higgs production and decay modes can be straightforwardly mapped onto this parameter space.   These parameters can in turn be easily calculated in any ultraviolet completion consistent with the assumptions of the framework, allowing one to quickly extract their effects on Higgs measurements.  For example, we showed how our framework allows one to easily map the NMSSM onto precision Higgs observables in electroweak gauge bosons and $\gamma \gamma$, without making use of parameter scans. 

The present work leaves a number of open questions for future study.  For example, one could consider how a determination of the mixing angles and loop parameters might be used to constrain the properties of the heavier scalar partners of the Higgs boson.  Concretely, if the mixing angles for the Higgs boson, $(\alpha,\gamma)$, can be identified by experiment, then unitarity constraints on the scalar mixing angles will in turn restrict the couplings of additional heavy Higgs bosons.  One interesting question is whether direct or indirect constraints in these theories will ultimately prove more powerful.  Another topic to study further is connecting our formalism for precision Higgs physics to other indirect constraints, {\it e.g.} from precision electroweak measurements \cite{Grojean:2013kd}, questions of vacuum stability, and viability of the new electroweak states as dark matter ({\em e.g.} as in \cite{Cheung:2012nb}).  

Precision Higgs physics will be an important probe for physics beyond the Standard Model.  Should the LHC experiments discover new physics in the Higgs sector, we have presented here a simple roadmap for understanding and parameterizing these effects beyond the SM.

{\em Acknowledgments:}  SDM and KZ are supported by NSF CAREER award PHY 1049896.  KZ is also supported by the DoE under contract de-sc0007859.

\end{document}